\documentclass[aps,twocolumn,superscriptaddress,floatfix,nofootinbib,longbibliography]{revtex4-1}
\usepackage{graphicx,amsmath,amsfonts,amssymb,amsthm,xr}
\usepackage{epsfig,amsmath,amssymb,color,dsfont,upgreek,physics, bm}
\usepackage{mathrsfs}
\usepackage{mathtools}
\usepackage{bbold}
\usepackage{comment}
\usepackage{float}
\usepackage[bookmarks=true,colorlinks,linkcolor=Blue,urlcolor=Blue,citecolor=Blue]{hyperref}
\usepackage[dvipsnames]{xcolor}
\usepackage{orcidlink}
\usepackage[caption=false]{subfig}
\usepackage{todonotes}

\usepackage{tikz}
\usetikzlibrary{arrows.meta, positioning}

\begin{document}
\title{Towards Interpretability of Neural Quantum States}

\author{Fabian Döschl\orcidlink{0009-0005-5067-004X}}
\email{Fa.Doeschl@physik.uni-muenchen.de}
\affiliation{Ludwig-Maximilians-University Munich, Theresienstr. 37, Munich D-80333, Germany}
\affiliation{Munich Center for Quantum Science and Technology, Schellingstr. 4, Munich D-80799, Germany}
\author{Annabelle Bohrdt\orcidlink{0000-0002-3339-5200}}
\affiliation{Ludwig-Maximilians-University Munich, Theresienstr. 37, Munich D-80333, Germany}
\affiliation{Munich Center for Quantum Science and Technology, Schellingstr. 4, Munich D-80799, Germany}

\date{\today}
\begin{abstract}
Neural quantum states (NQS) have emerged as a powerful variational ansatz for representing quantum many-body wave functions. Their internal mechanisms, however, remain poorly understood. We investigate the role of correlations for NQS-like quantum state representation by employing a correlation-based interpretable neural network architecture and then proving our observations using Boolean function theory. The correlator neural network demonstrates that, even for simple product states, up to all system-size correlation orders in the chosen computational basis are required to represent a quantum state faithfully. We explain these observations using Fourier expansion, which reveals the correlator basis as the effective basis of the internal NQS structure, the resulting necessity for high-order correlations that is supported by an entanglement bound that scales with the correlation order, consequences of linear dependencies in constrained Hilbert spaces for correlation requirements, and connections between spin basis rotations and the correlator basis. Furthermore, we analyze how neural networks achieve high correlation orders by increasing the magnitude of the network weights, which can be compensated by increasing the network depth. Lastly, we discuss how activation functions, network architectures, and choice of reference basis influence correlation requirements. Our results provide new insights and a better understanding of the internal structure and requirements of NQS, enabling a more systematic use of NQS in future research.
\end{abstract}

\maketitle

\section{Introduction}
Two-dimensional quantum systems pose one of the most interesting challenges for current quantum many-body physics, as they are central to our understanding of high-temperature superconductivity, effects in strongly correlated spin systems, and topological phases. However, studying these systems with analytical or numerical methods is notoriously difficult.

State-of-the-art methods, such as tensor network methods (TN) or quantum Monte Carlo (QMC) approaches are limited in their capability to find and represent quantum states in two-dimensional systems~\cite{Foulkes_2001,Becca_Sorella_2017,Pan_2024,Schollw_ck_2011}. Recently, Carleo and Troyer proposed a variational ansatz that uses the expressive power of neural networks to represent quantum states~\cite{Carleo_2017}. These so called neural quantum states (NQS) have shown remarkably good performance for many quantum systems~\cite{chen_2023,Chen_2025,Viteritti_2023,Lange_2024,schmitt_2025} and have proven capable of capturing quantum states with sign problem (limitation of QMC)~\cite{Giuliani_2023,Lange_2024_RNN,Doeschl_2025,Zakari_2025,Kufel_2025} or volume law entanglement (limitation of TN)~\cite{Gao_2017,Sharir_2022,Deng_Quantum_Ent_2017}. However, unlike TN and QMC, NQS operate as a black-box method, making their controllability and a general understanding of how they work and when they fail still elusive. In particular, it is unclear how the network’s internal structure is related to the correlations in the target quantum state.

Since the introduction of neural quantum states, there has been a significant interest in understanding their representational power~\cite{Gao_2017,Glasser_2018,Sharir_2022,barton_2025}, optimization properties~\cite{Carleo_2018,Kol_Namer_2024,chen_2023,Rende_2024_LinAl,Dash_2025,Moss_2025}, and practical limitations~\cite{Passetti_2023}. 
Despite the difficulty in analyzing and interpreting these non-linear variational functions, several studies have investigated the properties of neural quantum states analytically. A central result, extending beyond the study of NQS, is the proof that neural networks, given a potentially exponential number of parameters, can approximate any continuous function~\cite{Cybenko_1989,Hornik_1989,Hornik_1990}. Although this universality also applies to the NQS-like representation of arbitrary quantum state wave functions, it does not guarantee parameter efficiency. Nevertheless, in practice, we know from many numerical~\cite{hibatallah_2023,VariationalHibat,Pfau_2024,gu_2025,chen_2025_CNN,Zakari_D_2025} and analytical~\cite{Gao_2017,Deng_2017,Glasser_2018,Kaubruegger_2018,Lu_2019,yang_2024} examples that there exists, in many cases, an efficient representation of physically relevant quantum states. 

However, despite knowing about the promising properties of neural quantum states, it is often extremely challenging to fine-tune a network architecture for a specific quantum system such that we obtain the best results. A major hurdle for the optimization of NQS architectures is the fact that we often lack a clear understanding of the internal mechanisms of neural networks. A possibility to achieve a better understanding of the working principle is to draw connections between the neural network and the underlying correlations in the system. A step in this direction was taken in the article~\cite{Valenti_2022}, where weighted correlations (dominant terms in the Hamiltonian) were added to a restricted Boltzmann machine to improve convergence.

However, instead of improving a certain network architecture, we aim to reveal the required internal network structure for simulating quantum states. From the machine learning perspective, this was already done for phase and image classification tasks, where correlation based neural networks have been used to determine the important physical features for a successful characterization in terms of higher-order correlations~\cite{Miles_2021,Fischer_2022,Suresh_2024,cybinski_2024}. 
The crucial idea for these interpretable neural network architectures is to replace the standard non-linear activation function by a controlled expansion in higher-order correlations of the input, e.g. spin-spin correlations. Through regularization path analysis or tuning the maximal order of correlations considered as a hyperparameter, it is then possible to determine the most relevant correlation orders for a given \emph{classification} task.

In these cases, it was enough to use low correlation orders (up to fourth order) to distinguish different phases. Yet, these insights do not directly transfer to quantum states, raising the question of which correlations are important to represent a quantum state, rather than merely characterizing it. Since representation requires the full set of wave function coefficients, it may depend on correlations far beyond those relevant for phase detection.

In this paper, we systematically investigate which correlations are relevant to represent a quantum state. Using the real valued correlator transformer architecture~\cite{Suresh_2024}, a polynomial ansatz with controllable expansion order, we study how truncating the expansion affects the representational power for two paradigmatic models: the two dimensional Ising model and the toric code model. In combination with tools from Boolean function analysis, we identify conditions under which high correlation orders are unavoidable, even for simple product states, and when these requirements can be relaxed by Hilbert space truncations. This originates from the fact that the effective internal basis that the NQS uses to represent a quantum state, i.e. in which the connection between neural network parameters and wave function coefficient arises naturally, is not the reference spin (or Fock) basis, but the correlator basis. This framework allows us to derive consequences for entanglement scaling, and to relate these insights to general neural quantum states by comparing their Taylor expansion to the required polynomial structure.  Using the acquired knowledge, we discuss the differences between non-autoregressive and autoregressive neural networks and comment on commonly used network architectures that provide natural (dis)advantages from a Boolean function point of view.

The paper is structured as follows: In Section~\ref{NQS_definition} we introduce the NQS framework formally. Section~\ref{Observations} presents numerical results using the Correlator Quantum State (CQS), a polynomial architecture with controllable expansion order, for the 2D transverse field Ising model and the perturbed toric code, restricted to the physical sector. Section~\ref{Aofwbi} reviews Boolean function analysis that provides the analytical foundation that enables the explanation of our numerical observations, which is given in Section~\ref{Explaining_observations}. Section~\ref{General_Consequences} addresses the choice of computational basis, the entanglement bounds for fixed expansion orders, and the insights for general neural quantum states. Finally, in Section~\ref{Comment_on_architectures}, we draw conclusions about the desired NQS structure, enabling a structured optimization of its activation functions and architectures.

\section{Neural Quantum States}
\label{NQS_definition}
For an exact representation of a quantum state, one generally needs Hilbert space dimension $H_\mathrm{dim}$ many wave function coefficients to describe the system in terms of basis states,
\begin{equation}
\label{NQS:EQ1}
    |\Psi_\lambda \rangle = \sum_\sigma \psi_\mathrm{state}^\sigma | \sigma\rangle.
\end{equation}
Without loss of generality, we will mainly focus on spin-$1/2$ systems in this paper.
As storing and evaluating all $H_\mathrm{dim}$ coefficients is exponentially costly, one searches for a function that efficiently approximates these wave function coefficients $\psi_\mathrm{state} = \{\psi_\mathrm{state}^\sigma \}$. Neural quantum states (NQS) approach this by using a neural network to represent such a tunable function
$\psi_\theta(\sigma)$ which maps each basis state $\sigma $ to an output $A^\sigma_\theta$:
\begin{equation}
\label{NQS:EQ2}
    \psi_\theta: \{ -1,1\}^L \longrightarrow A_\theta \subseteq \mathbb{C} \quad \mathrm{with~} |A_\theta| \leq 2^L,
\end{equation}
where $A_\theta$ is the subset of $\mathbb{C}$ that contains all wave function coefficients $\{A^\sigma_\theta\}$. For a trained model, the normalized set $\frac{A_\theta}{\|A_\theta\|_2}$ is expected to approximate the wave function of the desired state $\frac{A_\theta}{\|A_\theta\|_2}\approx \psi_\mathrm{state}$. However, there is no guarantee that there exists a non exponentially scaling set of parameters $\theta$ such that the function $\psi_\theta(\sigma)$ represents the desired quantum state exactly or even approximately. 

\textit{Correlator Quantum States.---} 
The idea that any network architecture can be expanded using Taylor series (at least locally), makes a correlator neural network~\cite{Miles_2021,Suresh_2024,cybinski_2024} that simulates the Taylor expansion up to some order $n$, a valuable tool to acquire information about the required network structure. In our study, we analyze the structure by changing the maximal accessible expansion (correlation) order $n$. We define the highest expansion order $n$ by all ${L\choose n} $ correlations $\mathcal{X}_S(\vec{\sigma})$:
\begin{equation}
    \mathcal{X}_S(\vec{\sigma}) = \prod_{i \in S} \sigma_i,
\end{equation}
with $|S| = n$. Note that $S$ is a set of $n$ out of $L$ sites $i$ that specifies the product of spins $\sigma_i$ within a spin configuration $\vec{\sigma}$.

For clarity, when discussing correlations, we refer to products over spins within one sample $\sigma$, as opposed to expectation values of correlators such as $\langle\hat{S}_i^z \hat{S}_j^z\rangle$.

For a neural quantum state, the neural network maps a basis configuration $\sigma$ to a wave function coefficient $\psi_\theta(\vec{\sigma})$ through a combination of linear operations and applications of non-linear activation functions. The non-linear activation functions in principle enable contributions of arbitrarily high-orders of correlation functions between the inputs (e.g. between $\hat{\sigma}^z_i$ at different lattice sites).

To numerically investigate the dependence of NQS on correlations $\mathcal{X}_S(\vec{\sigma})$, we employ the correlator transformer architecture proposed in Suresh et al.~\cite{Suresh_2024}. A detailed description is provided in the Appendix~\ref{App:CoTra}.

The main idea of this transformer is to operate without standard non-linear activation functions, and to instead employ the non-linearity of higher-order correlations within a snapshot. These correlations are weighted and then used to evaluate the desired system. Based on the model's performance, one can infer the importance of different correlation orders.

In the context of NQS, the correlator quantum state (CQS) wave function is defined as:
\begin{equation}
    |\psi\rangle = \sum_{\vec{\sigma}}  \psi(\vec{\sigma}) |\vec{\sigma}\rangle\quad \mathrm{with} \quad \psi(\vec{\sigma})  = W^{\mathrm{pred}}\mathbb{X}(\vec{\sigma}) +\beta,
\end{equation} 
where $\mathbb{X}$ contains all correlation matrices $\mathbb{X} = [\bar{X}^1,\bar{X}^2,\dots,\bar{X}^i ,\dots,\bar{X}^n ]$. Each entry of $\bar{X}^i$ contains a weighted sum over all correlation orders $j$ of the same parity with $j\leq i$. To simplify the interpretation, we solely focus on real valued weights and wave function coefficients. 

In the following, we analyze the general structure of CQS based on the mathematical framework of Boolean functions and investigate the underlying requirements on network architectures to represent quantum states.

\section{Quantum state characterization vs. representation}
\label{Observations}
It is well-established that conventional phases of matter can be probed and characterized using local observables. For instance, in Ising systems, magnetization serves as a key local observable that reveals information about the underlying phase.

This concept of characterizing a quantum state based on local information can also be found in machine learning. Following the ideas of interpretability, many articles~\cite{Miles_2021,cybinski_2024,Suresh_2024,Fischer_2022} showed an efficient phase detection by considering a small number of low-order correlations within the system.

Similarly, one can ask a network which and how many correlations it needs to represent a quantum state. At first glance, one might assume that the highest order contributing when representing a quantum state is the same as the one required to characterize it. Although these two tasks appear related, their required structure is inherently different.

Even when performing a Taylor expansion of the neural network architecture around a point $\vec{p}$:
\begin{equation}
    \psi_\theta(\vec{\sigma}) = \sum_{n=0}^\infty \frac{\psi_\theta^{(n)}(\vec{p})}{n!} (\vec{\sigma}-\vec{p})^n,
\end{equation}
one might also expect main contributions from lower correlation orders. Here, $\psi_\theta(\vec{\sigma}) $ can be any arbitrary network architecture with weights $\theta$ and a converging Taylor series. The $1/n!$ in the Taylor series suppresses higher-order terms by construction, as required for convergence. Therefore, one might conclude from working NQS examples that only lower-order correlations play a significant role.
In the following, we will show that this is not necessarily the case. We first show numerical results obtained by using a correlator transformer architecture, and then explain the findings based on the mathematical framework of Boolean functions.\\

\begin{figure*}[t]
  \centering
  \includegraphics{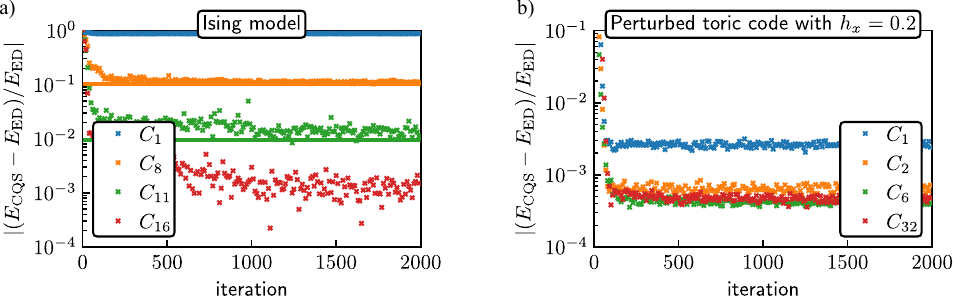}
  \caption{Convergence of the correlator transformer (square lattice of linear size $L = 4$) applied to (a) the Ising model and (b) the perturbed toric code. $C_n$ defines a network that uses all correlations up to order $n$. The horizontal lines shown in (a) are the (exact) lowest possible energies for the respective maximal correlation order. The exact energies in the correlator basis were calculated using Eq.~\eqref{Eq:Ham_corr}.}
  \label{Fig:Fig_order}
\end{figure*}

\textit{Numerical observations.---} In the following, we apply the real valued correlator transformer to learn the ground state of the Ising model and the perturbed toric code. To probe the representational requirements of CQS for these systems, we vary the highest accessible correlation order $n$. The CQS, denoted $C_n$, with maximal order $n$ then contains all possible correlations up to and including order $n$, which we informally write as:
\begin{equation}
    C_n = \sum^n_{i=0}\sum_{S\subseteq[L]}\mathcal{X}_{S} (\vec{\sigma})\delta_{|S|,i},
\end{equation}where $[L]$ denotes the set of all possible combinations of sites in the system. The resulting convergence of the energy is shown in Fig.~\ref{Fig:Fig_order}\,a) for the Ising model and b) for the perturbed toric code, respectively.

First, we show that our observations contradict naive expectation; then we review tools from Boolean function analysis (Section~\ref{Aofwbi}) that allow us to explain our results in Section~\ref{Explaining_observations}. \\

\textit{Ising model.---} We start by applying the correlator transformer NQS (CQS) to the Ising model:
\begin{equation}
    \hat{H}_\mathrm{Ising} =  \sum_{ \langle i,j \rangle} \hat{\sigma}^z_i \hat{\sigma}^z_j,
\end{equation}
where $\sigma^\alpha$ are the standard Pauli operators. We define our system in the $\sigma^z$ basis.
This Hamiltonian has two-body interactions and the emergent ground state phases are fully characterized by first- and second-order correlations. Even with machine learning, these correlations are sufficient to characterize the phases~\cite{cybinski_2024,Suresh_2024}.

As shown in Fig.~\ref{Fig:Fig_order}\,a), which plots the energy convergence of different CQS $C_n$, the correlator transformer architecture requires access to all correlation orders to obtain the best results. Therefore, a quantum many-body wave function does not generally fall off at high correlation orders. 
Note that poor performance of our CQS for lower correlation orders cannot be ruled out a priori, as the ansatz may be unable to represent the state or to find the optimal parameters. To exclude convergence or expressivity issues of the CQS, we also computed the exact lowest energy for the respective maximal correlation order $C_n$. The minimal energies for a given expansion order are displayed as horizontal lines in Fig.~\ref{Fig:Fig_order}\,a). The derivation of the Hamiltonian in the correlator basis is given in Section~\ref{MathFramework}.

While $C_1 = \mathcal{X}_\emptyset + \mathcal{X}_{|S| = 1} =\mathrm{bias} + \mathrm{order~} 1 $ completely fails to converge to the antiferromagnetic (AFM) ground state, $C_8$, which includes correlations up to $8^\mathrm{th}$ order, achieves better results. However, the relative error remains at approximately $\epsilon \approx 10\%$. A further increase in the order of correlations significantly improves the results. With $C_{11}$, the relative error is reduced to approximately $\epsilon \approx \mathcal{O}(10^{-2})$. However, the best performance, $\epsilon \approx \mathcal{O}(10^{-3})$, is achieved only when all correlation orders are included ($C_L =C_{16}$).
From the naive perspective, this is unexpected, as even $C_8$ and $C_{11}$, that contain all correlations up to order $8$, $11$ respectively, still fail to fully represent the antiferromagnetic ground state. 

We emphasize the importance of this finding: The AFM state is a product state - the simplest form of a quantum state. While it is trivial to encode this state as a Matrix Product State with bond dimension 1, CQS in the $|\vec{\sigma}_z\rangle$ basis requires access to all $L$ correlation orders in the system. Note that this does not imply that product states are fundamentally difficult for CQS to represent, rather, it shows that the required polynomial expansion order can scale linearly with system size, even for simple states.

In Section~\ref{Explaining_observations} we explain our findings based on a mathematical framework. For instance, it can be shown analytically that the correlation order $16$ is (for the AFM ground state) a significant contribution to the wave function.\\

\textit{Toric code model.---} We now turn to the perturbed toric code model, which is, for our study, particularly interesting, as it exhibits topological order due to four-body interactions. The corresponding Hamiltonian is defined by:
\begin{equation}
 \hat{H}_\mathrm{TC}  = -\sum_{\mathrm{j}}\prod_{i \in +_j} \hat{\sigma}_i^x - \sum_{\mathrm{j}}\prod_{i \in \square_j} \hat{\sigma}_i^z + h_x \sum_i \hat{\sigma}^x_i.
\end{equation}
Again, $\sigma^\alpha$ are the usual Pauli operators. We define the spins in the system as the connecting links between sites, such that the total Hilbert space dimension is given by $2^{2(L_x\times L_y)}$. For the CQS simulation, we choose to constrain the Hilbert space to the physical sector where Gauss's law:
\begin{equation}
    G_i = \prod_{j \in +_i} \hat{\sigma}_i^x \overset{!}{=} 1
\end{equation}
is always fulfilled, thus reducing the effective Hilbert space to $2^{(L_x\times L_y)+1}$~\cite{Luo_2021,Sheffer_2025}. This is achieved by choosing the $\sigma^x$ basis as the computational basis, which allows us to restrict the Monte Carlo updates to states from the physical sector. \\

The perturbed toric code is a Hamiltonian with a topological phase, which makes it, in principle, difficult to characterize the relevant information for a given phase or wave function. For phase detection tasks, it was shown that $4^\mathrm{th}$ order correlations are sufficient to achieve a successful classification~\cite{Suresh_2024}. Similar to the Ising model, one can show that an CQS requires all correlations in the system if one works in the full Hilbert space (see Section~\ref{Explaining_observations}). However, in the considered case with a Hilbert space truncation, it is at first unclear which and how many correlations are required for a full description.

Note that if we enforce Gauss's law, the limiting case of the pure toric code model ($h_x = 0$) has a trivial solution, which is given by a constant $\psi^{h_x = 0}_\mathrm{res}(\vec{\sigma}) = C_0$.

In Fig.~\ref{Fig:Fig_order}b), we show the convergence of the correlator transformer model for the perturbed toric code with $h_x = 0.2$. One immediately notices that the perturbed toric code, despite the topologically ordered phase, is significantly easier to learn than the AFM ground state. All evaluated correlator transformers achieve a reasonable error of $\epsilon \lesssim \mathcal{O}(10^{-3})$. This can be attributed to the fact that the system at $h_x = 0.2$ remains in the topologically ordered phase, so that only minor deviations from the unperturbed ground state $\psi^{h_x = 0}_\mathrm{res}(\vec{\sigma}) = C_0$ are expected.

While $C_1$, which contains first-order correlations and a bias, already achieves a reasonable error of $\epsilon \approx \mathcal{O}(10^{-3})$, the results can still be improved by providing access to higher-order correlations. However, already when using $C_6$, an optimization limit is reached, where we do not observe any further improvement when increasing the number of accessible correlation orders. Even when access to all correlations is allowed in $C_{32}$, we do not observe a major difference in the convergence or the energy errors.

Although a good performance of the CQS for the slightly perturbed toric code with enforced Gauss's law is generally expected, the stagnation observed for $C_{\geq6}$ is surprising, given that the number of correlations accessible to the $C_{32}$ CQS is several orders higher than those available to the $C_{6}$ CQS (compare with Eq.~\eqref{Eq:Num_Corrs}). In Section~\ref{Explaining_observations}, we explain the reduced number of required correlations using the decision tree formalism described in Section~\ref{DecisionTreeForm}. In particular, we show explicitly how the linear dependencies induced by Gauss's law reduce the correlation orders needed to represent the ground state.\\

\textit{Remark.---} The best energies that we obtained in the training are limited by the representational power of our network architecture. However, the correlator transformer was built to yield interpretability at the cost of reduced complexity. Together with the limited number of parameters, which we set to $ \#\mathrm{params} \approx \mathcal{O}(2\times10^4)$ for all CQS simulations, we do not expect a perfect convergence.

\section{Analysis of functions with binary inputs}
\label{Aofwbi}
As we numerically observed in the previous section, neural quantum states cannot behave the same as typical machine learning tasks, where mostly low correlation orders are relevant~\cite{Miles_2021,Suresh_2024,cybinski_2024}. In the following, we review the Boolean function formalism that reveals the correlator basis as the internal, input independent basis of NQS-like methods (Section~\ref{MathFramework}). Thereafter, we discuss the decision tree formalism that provides a constructive framework for systems with truncated Hilbert spaces, and thereby explains the potential relaxation of correlation order requirements (Section~\ref{DecisionTreeForm}). Finally, we provide a more familiar picture of why a broad correlation order spectrum is needed, by looking at a set of equations (Section~\ref{Sec_intuitiveFE}).

\subsection{Boolean function analysis}
\label{MathFramework}
While neural quantum states are often perceived as a purely ``black box'' technique, we show that a much better understanding of the working principle can be achieved when studying the underlying structure. To do so, we follow the mathematical framework of Boolean functions~\cite{ODonnell_2014}. Note that there is an extension of this framework to higher local Hilbert space dimensions and to one-hot encoding. We comment on these generalizations in Appendix~\ref{App:Generalization}.\\

The analysis of real valued Boolean functions $\psi:\{-1,1\}^L \rightarrow \mathbb{R}$ is based on studying their Fourier expansion. The Fourier expansion can be readily derived by introducing an indicator polynomial:
\begin{equation}
    1_{\vec{\alpha}}(\vec{\sigma}) =  \frac{(1+\alpha_1 \sigma_1)}{2} ... \frac{(1+\alpha_L \sigma_L)}{2} = \left\{ \begin{matrix}
        1, \; \vec{\sigma} = \vec{a}, \\
        0, \; \vec{\sigma} \neq \vec{a}.
    \end{matrix} \right.
\end{equation}
This indicator function is a multilinear polynomial that includes correlations up to $L^\mathrm{th}$ order. Consequently, any wave function of a finite-size spin system has a polynomial representation of the form:
\begin{equation}
    \psi(\vec{\sigma}) = \sum_{\vec{\alpha} \in \{-1,1\}^L} \psi(\vec{\alpha} ) 1_{\vec{\alpha}} (\vec{\sigma})
\end{equation}
Rewriting this polynomial representation in terms of all possible (overlapping) subsets $S$ of $[L] = [\sigma_1,\sigma_2,...,\sigma_L]$ yields the Fourier representation of $\psi(\vec{\sigma})$:
\begin{equation}
\label{Eq:WaveFunc}
    \psi(\vec{\sigma}) = \sum_{S\subseteq [L]}   \langle \vec{\sigma}|\mathcal{X_S}\rangle  \langle \mathcal{X_S}|\psi\rangle =\sum_{S\subseteq [L]} \bar{\psi}(S) \mathcal{X}_S(\vec{\sigma}),
\end{equation}
where the corresponding Fourier coefficients are defined as:
\begin{equation}
\label{Eq:Fouriercoeff}
    \bar{\psi}(S) =\langle  \mathcal{X_S} |\psi \rangle = \frac{1}{2^L}\sum_{\vec{\sigma} \in \{-1,1\}^L} \psi(\vec{\sigma}) \mathcal{X}_S(\vec{\sigma}),
\end{equation}
and the monomial $\mathcal{X}_S(\vec{\sigma})$ corresponding to $S$ is given by:
\begin{equation}
\label{Eq:Basis}
    \mathcal{X}_S(\vec{\sigma}) = \langle \vec{\sigma}|\mathcal{X_S}\rangle = \prod_{i\in S} \sigma_i,
\end{equation}
with $\mathcal{X}_\emptyset(\vec{\sigma}) = 1$. This construction yields in total $2^L$ monomials that form an orthonormal basis for the $H_\mathrm{dim} = 2^L$ dimensional Hilbert space~\cite{ODonnell_2014}:
\begin{equation}
\label{Eq:Orthogonal}
    \langle \mathcal{X}_S|\mathcal{X}_T\rangle=\frac{1}{2^L} \sum_{\vec{\sigma} \in \{-1,1\}^L} \mathcal{X}_S(\vec{\sigma}) \mathcal{X}_T(\vec{\sigma}) = \left\{ 
    \begin{matrix}
        1, \; S = T, \\
        0, \; S\neq T.
        \end{matrix}
        \right. 
\end{equation}
Note that the $k^\mathrm{th}$ correlation order (all $S$ of the same degree: $|S| = k$) contributes $\frac{L!}{k!(L-k)!}$ monomials to the basis. Thus, a CQS $C_n$ with correlations up to $n^\mathrm{th}$ order has access to $N(C_n)$ basis vectors:
\begin{equation}
\label{Eq:Num_Corrs}
    N(C_n) =\sum_{k = 0}^n \frac{L!}{k!(L-k)!} = 
        \sum_{k = 0}^n {L\choose k} \overset{n=L}{=} 2^{L}.
\end{equation}
Consequently, ground states (GS) of Hamiltonians:
\begin{equation}
\label{Eq:Ham_corr}
    \hat{H} = \sum_{S,S'\subseteq [L]} \underbrace{\left( \frac{1}{2^L}\sum_{\vec{\sigma}, \vec{\sigma}'} \langle \mathcal{X_S}| \vec{\sigma}\rangle h_{\vec{\sigma},\vec{\sigma}'}\langle \vec{\sigma}'|\mathcal{X_{S'}}\rangle \right) }_{h_{S,S'}} |\mathcal{X_S}\rangle\langle \mathcal{X_{S'}}| ,
\end{equation}
with relevant contributions $h_{R,S'}$ in higher-order correlations $|R| >n $, such that $\bar{\psi}_\mathrm{GS}(R){\neq}0 $, cannot be fully represented by the CQS $C_n$ for $n<|R|\leq L$. Here, $h_{\vec{\sigma},\vec{\sigma}'}$ is the usual matrix element $h_{\vec{\sigma},\vec{\sigma}'} = \frac{1}{2^L} \langle \vec{\sigma}|\hat{H}|\vec{\sigma}'\rangle$. Note that due to the normalization of $\langle \mathcal{X}_S|\mathcal{X}_T\rangle = \delta_{S,T}$ we used the convention $\langle \vec{\sigma}|\vec{\sigma}'\rangle = 2^L\delta_{\vec{\sigma},\vec{\sigma}'}$ and $\mathbb{1} = \frac{1}{2^L}\sum_{\vec{\sigma} \in \{-1,1\}^L}|\vec{\sigma}\rangle \langle \vec{\sigma}|$.

\textit{Remark.---} We want to stress the importance of this formalism for neural quantum states: Eq.~\eqref{Eq:WaveFunc} decouples the influence of the network weights, included in $\bar{\psi}(S)$, from the input configuration, given in $\mathcal{X}_S(\vec{\sigma})$. In other words, the machine learning algorithm approximates the quantum state in the correlator basis $|\mathcal{X}_S \rangle$ and rotates it into the desired spin basis:
\begin{equation}
|\psi \rangle = \sum_{S\subseteq [L]} \bar{\psi}(S) | \mathcal{X}_S \rangle \Rightarrow \sum_{\vec{\sigma} \in \{-1,1\}^L}\sum_{S\subseteq [L]} \bar{\psi}(S) \langle\vec{\sigma}| \mathcal{X}_S \rangle |\vec{\sigma}\rangle
\end{equation}
Therefore, we can think of the correlator basis $|\mathcal{X}_S \rangle$ as the internal basis that neural quantum states use for quantum state representation. Consequently, operations that are simple in the spin basis, such as setting wave function coefficients $\psi(\vec{\sigma})$ to zero for selected configurations, can translate into constraints on $\bar{\psi}(S)$ that may require cancelations among multiple potentially non-trivial coefficients. This does not rule out efficient neural network representations, but shows that such efficiency cannot, in general, be inferred from the simplicity or sparsity in the spin basis for an arbitrary NQS ansatz.


\subsection{Decision tree}

\label{DecisionTreeForm}
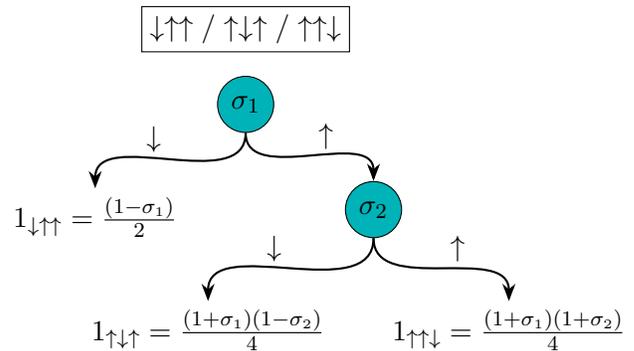
\begin{figure}[t]
  \centering
\begin{tikzpicture}[>=Stealth, every node/.style={font=\fontsize{11}{13}\selectfont}]
  \node[rectangle,draw, fill=white] (states) at (0,1) {$\downarrow \uparrow \uparrow / \uparrow \downarrow \uparrow/\uparrow \uparrow\downarrow$};
  \node[circle, draw, fill=BlueGreen] (sigma1) at (0,0) {$\sigma_1$};
  \node[circle, draw, fill=BlueGreen] (sigma2) at (1.7,-1.4) {$\sigma_2$};
  \node (leaf1) at (-2,-1.5) {$1_{\downarrow \uparrow \uparrow} = \frac{(1-\sigma_1)}{2}$};
  \node (leaf2) at (-0.5,-3.) {$1_{\uparrow \downarrow \uparrow} = \frac{(1+\sigma_1)(1-\sigma_2)}{4}$};
  \node (leaf3) at (3.5,-3.) {$1_{\uparrow \uparrow\downarrow } = \frac{(1+\sigma_1)(1+\sigma_2)}{4}$};

  \draw[->, thick, black] (sigma1) to[out=270, in=90] node[pos=0.5, right, yshift=8pt] {$\uparrow$} (sigma2) ;
  \draw[->, thick, black] (sigma1) to[out=270, in=90] node[pos=0.5, left, yshift=8pt] {$\downarrow$} (leaf1);

  \draw[->, thick, black] (sigma2) to[out=270, in=90] node[pos=0.5, right, yshift=8pt] {$\uparrow$} (leaf3) ;
  \draw[->, thick, black] (sigma2) to[out=270, in=90] node[pos=0.5, left, yshift=8pt] {$\downarrow$} (leaf2) ;

\end{tikzpicture}
\caption{Decision tree for a three spin model with a fixed magnetization of $\langle \sum_i \hat{\sigma}^i_z\rangle = 1$.}
\label{Fig:DecisionTree}
\end{figure}
Although the Fourier expansion described above provides a good starting point for the analysis, it becomes challenging to understand the behavior of NQS in systems with artificially reduced Hilbert spaces, such as fixed magnetization, Gauss's law, or particle conservation. 
To address this, one can introduce a decision tree representation (see Fig.~\ref{Fig:DecisionTree}). Similar to autoregressive NQS models~\cite{Hibat_Allah_2020,AnnealingHibat_Allah}, this approach constructs computation paths through the lattice and sequentially checks for allowed (disallowed) local spin configurations. Following this, the restricted wave function $\psi_\mathrm{res}$ can be written as:
\begin{equation}
    \psi_\mathrm{res}(\vec{\sigma}) = \sum_{\vec{p} \in T} \psi(\vec{p}) 1^\mathrm{res}_{\vec{p}}(\vec{\sigma}),
\end{equation}
where $\vec{p}$ denotes a valid path in the decision tree $T$. Again, $1^\mathrm{res}_{\vec{p}}(\vec{\sigma})$ is an indicator polynomial built from $\frac{(1+p_i \sigma_i)}{2}$ elements. Once a branch in the tree becomes trivial (e.g. all particles are distributed), this polynomial ends.

In the context of the Fourier (or Taylor) expansion, the restriction of the Hilbert space reduces the maximal correlation order in the Fourier representations to the maximal length of a branch $\max|\vec{p}|$ (compare with Fig.~\ref{Fig:DecisionTree}). Importantly, due to the truncation of the Hilbert space $\mathcal{H}_\mathrm{res}$, many of the monomials $\mathcal{X}_S$ become linearly dependent, which implies that the Fourier basis loses (at least partially) its orthogonality: $\max|\vec{p}| \geq \log_2\dim(\mathcal{H}^\mathrm{res}_\mathrm{space})$.

So far, we have considered only one single decision tree. However, decision trees are not unique. Therefore, a neural network can take the form of a sum over all possible decision trees $\mathbb{T} = [T_1,T_2,...]$ for each configuration and select the set of paths with minimal depth to the desired state:
\begin{equation}
    \psi_\mathrm{res}(\vec{\sigma}) = \sum_{\vec{\alpha} \in \mathcal{H}_\mathrm{res}} \psi(\vec{\alpha}) \sum_{\vec{p} \in \mathbb{T}} c_{\vec{\alpha},\vec{p}} 1^\mathrm{res}_{\vec{p}}(\vec{\sigma}),
\end{equation}
where $c_{\alpha,\vec{p}}$ has to fulfill:
\begin{equation}
    \sum_{\vec{p} \in \mathbb{T}} c_{\vec{\alpha},\vec{p}} 1^\mathrm{res}_{\vec{p}}(\vec{\sigma})  = \left\{ \begin{matrix}
        1, \; \vec{\sigma} = \vec{a}, \\
        0, \; \vec{\sigma} \neq \vec{a}.
    \end{matrix} \right.
\end{equation}

For the three-spin example with a fixed magnetization of $\langle \sum_i \hat{\sigma}^i_z\rangle = 1$, shown in Fig.~\ref{Fig:DecisionTree}, this is equivalent to choosing:
\begin{equation}
    1_{\downarrow \uparrow \uparrow} = \frac{(1-\sigma_1)}{2}, \, 1_{\uparrow \downarrow \uparrow} = \frac{(1-\sigma_2)}{2},\,1_{\uparrow \uparrow \downarrow} = \frac{(1-\sigma_3)}{2}.
\end{equation}
Note that $\sigma_3$ is constrained by $\sigma_3 = 1 - (\sigma_1+\sigma_2)$.

Importantly, not every truncation of the Hilbertspace or symmetry of the wave function reduces the maximal required expansion order. For example, using the spin flip symmetry:
\begin{equation}
    \psi_\mathrm{eff}(\sigma) = \frac{1}{2}[\psi(\sigma) + \psi(-\sigma)],
\end{equation}
removes all odd correlation orders, but leaves the even ones unchanged. Therefore, a reduced number of independent amplitudes does not by itself imply a reduced maximal correlation order. The decision tree formalism is a constructive framework that allows a better understanding for how linear dependencies from restricted Hilbert spaces influence the correlation requirements of neural networks. The minimal required correlation order for an exact representation of a quantum state is given by the maximal branch depth of the optimal set of decision trees, which can be significantly smaller than the system size.

\textit{Remark.---}  
In classical machine learning tasks, such as image recognition, the training data constitutes only a small fraction of the full configuration space. Consequently, linear dependencies arise artificially due to sampling sparsity, which artificially reduces the correlation requirements of the respective tasks.

\subsection{Intuitive approach to Fourier expansion}
\label{Sec_intuitiveFE}
In the Sections~\ref{MathFramework} and~\ref{DecisionTreeForm}, we showed formally that every function $\psi:\{-1,1\}^L \rightarrow \mathbb{R}$ can be written as a Fourier expansion~\cite{ODonnell_2014}. Here, we provide a more intuitive picture of what this means for neural quantum states.

Writing out the Fourier expansion of a wave function $\psi(\vec{\sigma})$ explicitly reveals its structure:
\begin{equation}
\label{Eq:MPR_Eq_X}
\begin{split}
      \psi (\vec{\sigma}) = & W + \sum_i A_i  \sigma_i +\sum_{j<i}B_{ij} \sigma_i \sigma_j + ... \\&+F\sigma_1\sigma_2\sigma_3... \sigma_L,
\end{split}
\end{equation}
where each variable ($W, A_i,B_{ij}, ...,F$) correspond to a specific Fourier coefficient $\bar{\psi}(S)$. Notice the similarity to a Taylor expansion. Writing the wave function coefficient $\psi (\vec{\sigma})$ for each basis state yields a set of $2^L$ equations:
\begin{equation}
\label{Eq:SetOfEquations}
    \begin{array}{c|c}
    \mathrm{state} &  \psi(\sigma)   \\
    \hline
        |\uparrow\uparrow\uparrow...\uparrow\rangle &   A_1+A_2... + B_{12}+B_{13}  ...+F... + W     \\
        |\downarrow\uparrow\uparrow...\uparrow\rangle &   A_1-A_2... - B_{12}-B_{13}   ...-F... + W        \\
        \vdots&\vdots\\
        |\uparrow\downarrow\downarrow...\downarrow\rangle &   A_1-A_2... - B_{12}-B_{13}   ...\mp F... + W        \\
        |\downarrow\downarrow\downarrow...\downarrow\rangle&   -A_1-A_2... + B_{12}+B_{13}   ...\pm F... + W
    \end{array}
\end{equation}
The signs in front of the coefficients reflect the products of spins in that state. Importantly, the parity of the prefactor of F, $\sigma_1\sigma_2\sigma_3... \sigma_L = \pm1$, changes with the number of lattice sites. 

The task we aim to achieve with neural quantum states is solving these $2^L$ linearly independent equations, which generally requires all $2^L$ Fourier coefficients $A_i,B_{ij},...$, so we cannot arbitrarily assume that some coefficients are negligible. Note that setting coefficients of higher-order correlations to zero is equivalent to overdetermining the system, which can, depending on the quantum state, prevent an exact solution and make approximations difficult, if not impossible.

From this perspective, the Fourier representation provides a natural language to understand the requirements for neural quantum states. The network must be able to generate the set of correlations needed to represent the wave function. Whether this requires all orders or only a truncated subset depends on both the quantum state and the choice of computational basis, as we discuss in later sections.

\section{Explaining observations based on mathematical framework}
\label{Explaining_observations}
Based on the mathematical framework introduced in the Sections~\ref{MathFramework} and~\ref{DecisionTreeForm}, we now analyze the numerical observations made in Section~\ref{Observations}. We begin with the Ising model and thereafter turn to the perturbed toric code.

\textit{Ising model.---} In our numerical experiments, we found that all correlation orders are required to obtain a good ground state representation for a classical Ising model. At first glance, this may seem counterintuitive considering that the characterization of an antiferromagnetic state can be achieved by local measurements. However, with the framework discussed in Section~\ref{MathFramework}, we can now prove that the highest-order Fourier coefficient $\bar{\psi}(L)$ is a relevant contribution to the ground state. 

Note that we restrict this explanation to positive wave function coefficients $\psi(\sigma) \geq 0$ with an even number of spins. Depending on the parity of the number of lattice sites and the sign structure of the ground state, the highest contributing correlation order can shift to $L-1$ such that $\bar{\psi}(L-1) \geq \bar{\psi}(S)$).

Starting from Eq.~\eqref{Eq:Fouriercoeff}, we see that the number of down spins in the subset $S$ of the contributing basis states determines the magnitude of the Fourier coefficient. For an antiferromagnetic (AFM) state we obtain:
\begin{equation}
\label{Eq:AFM_State}
\begin{split}
           \bar{\psi}_\mathrm{AFM}(S) =&\frac{1}{2^L}\sum_{\vec{\sigma} \in \{-1,1\}^L} \psi_\mathrm{AFM}(\vec{\sigma})\prod_{i\in S} \sigma_i \\ =& \frac{1}{2^L} \left[ \psi_\mathrm{A} \prod_{i\in S} \sigma^{\mathrm{A}}_i + \psi_\mathrm{B} \prod_{i\in S}\sigma^{\mathrm{B}}_i\right] 
\end{split}
\end{equation}
Here, we used A and B for the two possible AFM states. In the considered $4 \times 4$ Ising model the AFM ground states contain an even number of down spins, implying that the two corresponding highest order monomials are $\mathcal{X}_L(\mathrm{even \downarrow})  = 1$. Plugging this into Eq.~\eqref{Eq:AFM_State} we obtain:
\begin{equation}
        \bar{\psi}(L) =\frac{1}{2^L} ( \psi_A +\psi_B) \geq \bar{\psi}(S),
\end{equation}
for any $S \subseteq[L]$, which makes the highest correlation order (for positive wave function coefficients) a significant contribution to the wave function. Therefore, the correlator transformer that was used for the Ising model in Section~\ref{Observations} with $C_8$ and $C_{11}$ cannot represent the full ground state although it has access to approximately $60\%$ and $96\%$ of all correlations (compare with Eq.~\eqref{Eq:Num_Corrs}), respectively.

Note that the general form of the Fourier coefficients for an AFM state can be derived using Eq.~\eqref{Eq:AFM_State}:
\begin{equation}
  \bar{\psi}(S) =    \left\{ 
    \begin{matrix}
        \pm \frac{1}{2^L} \left[ \psi_\mathrm{A} + \psi_\mathrm{B} \right]  & \mathrm{for~} |S| = \mathrm{even} , \\
        \pm \frac{1}{2^L} \left[ \psi_\mathrm{A} - \psi_\mathrm{B} \right] & \mathrm{otherwise},
        \end{matrix}
        \right.
\end{equation}
which shows that higher-order correlations are as relevant as lower correlation orders of the same parity. The large number of non-zero coefficients $\bar{\psi}(S)$ arises because the network must set most wave function coefficients $\psi_\mathrm{AFM}(\vec{\sigma})$ to zero.

\textit{Remark.---} This analysis can be readily extended to other quantum states constructed from basis states $\ket{\vec{\sigma}}$ with an even number of down spins. The toric code ground state in the full Hilbert space is an example of a more complex quantum state that shows the same behavior $\bar{\psi}_{\mathrm{TC}}(L) \geq \bar{\psi}_{\mathrm{TC}}(S)$.

\textit{Toric code model.---}
\begin{figure}[t]
    \centering
    \includegraphics{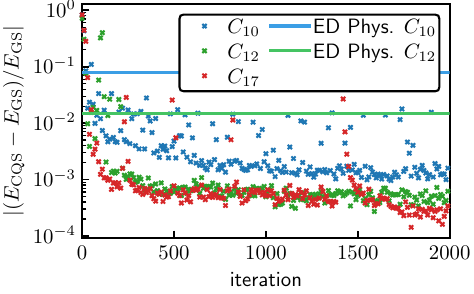}
    \caption{Convergence of the CQS (access to all sites) for the perturbed toric code ($4\times4$) model with enforced Gauss's law ($h_x = 1$ and star and plaquette term set to zero). We compare the CQS to exact diagonalization results (horizontal lines) in the correlation basis with correlations up to order $10$ (blue) and $12$ (green).
   }
    \label{Fig:TCvsED}
\end{figure}
The numerical simulation of the perturbed toric code model with enforced Gauss's law (compare with Fig.~\ref{Fig:Fig_order}\,b)) revealed that only a limited number of correlation orders played a significant role in obtaining the best possible ground state energy.

The Hilbert space of the toric code model restricted to the physical sector has dimension $2^{L_x\times L_y +1}$~\cite{Luo_2021,Sheffer_2025}, which means the system of equations is underdetermined (compare with Section~\ref{Sec_intuitiveFE}). Following the spanning tree formalism (see Appendix~\ref{App:SpanningTree}), an orthogonal basis can be constructed from $L_x\times L_y +1$ linearly independent spins. These spins uniquely determine the remaining $L_x \times L_y -1$ spins, fully specifying the state within the physical subspace. Therefore, using a decision tree (as described in Section~\ref{DecisionTreeForm}) that includes precisely these $L_x\times L_y +1$ spins, is sufficient for an exact representation of the physical sector. However, this construction requires capturing correlations up to order $L_x\times L_y +1$. Despite this, the results shown in Fig.\ref{Fig:Fig_order}\,b) indicate that lower-order correlations already suffice for an accurate approximation of the ground state $\psi^{h_x = 0.2}_\mathrm{res}$. 

One might argue that higher-order correlations are not essential in this regime, especially since $\psi^{h_x = 0.2}_\mathrm{res}$ remains close to the uniform $h_x = 0$ ground state $\psi^{h_x = 0.0}_\mathrm{res} = C_0$. To investigate this matter, we now consider the large $h_x$ limit. In the previous analysis it was shown that representing the ferromagnetic ground state $| ++...+\rangle$ in the $\sigma^x$ basis (large $h_x$ limit) requires all ($L_x\times L_y +1$) correlation orders in the reduced Hilbert space.

Fig.~\ref{Fig:TCvsED} shows the convergence of the CQS (with access to all sites) for various maximal correlation orders in the large $h_x$ limit. Furthermore, we compare the CQS energy to exact results obtained by diagonalizing the Hamiltonian in the orthogonal correlator basis (up to some correlation order) of the $L_x\times L_y +1$ linearly independent spins. Similarly to the behavior observed in Fig.~\ref{Fig:Fig_order}\,b), we notice that the results, beyond a certain correlation order,  do not improve further. All tested CQS obtain a minimal relative error of $\epsilon \approx \mathcal{O}(10^{-3})$. Furthermore, considering the same maximal correlation order, the CQS surpasses the exact results by several magnitudes. Here, the sole advantage of the CQS is the additional information about the remaining $L_x\times L_y -1$ spins. Consequently, we can conclude that the information about the predetermined $L_x\times L_y -1$ spins has to contain higher-order correlations of the $L_x\times L_y +1$ linear independent spins so that accurate results can be obtained even when not considering all $L_x\times L_y +1$ correlation orders. 

This is equivalent to saying that the spanning tree and therefore the decision tree in this toric code model is not unique. While we chose a specific decision tree for the exact diagonalization with a maximum correlation order of $L_x\times L_y +1$, we allowed the CQS to choose the shortest and most efficient path.

\textit{Remark.---} We emphasize that the exact diagonalization (compare with Fig.~\ref{Fig:TCvsED}) only uses states from the physical sector and is, therefore, still feasible for 32 spins.

\section{General consequences of the internal NQS basis}
\label{General_Consequences}
The Boolean function framework, which identifies the correlator basis as the internal NQS basis (Section~\ref{Aofwbi}), leads to several conclusions about the correlation properties required to represent a given quantum state. First, we establish the equivalence of the correlation transformation and the spin basis rotation, showing how the choice of computational basis influences the correlation structure of the target state (Section~\ref{ChoosingTheComputationalBasis}). Second, we derive entanglement bounds for CQS, demonstrating that a fixed expansion order can at most capture logarithmic entanglement scaling, whereas states with area or volume law entanglement require a CQS with correspondingly quasi area or volume scaling expansion order (Section~\ref{Consequences:Entanglement}). Finally, we use these insights to draw conclusions for general NQS by comparing their Taylor expansion with the polynomial description of a quantum state given by the Boolean function framework. This comparison reveals a bias in the exploration of Hilbert space and clarifies how the required correlation order depends on architectural details of the network, as discussed in Section~\ref{Sec:CompTaylor}.

Note, in the Appendix~\ref{App:relationSpinmomentsandlogpsi}, we address the relation between $\psi(\sigma)$, $p(\sigma) = |\psi(\sigma)|^2$, and $\log\psi(\sigma)$. Their Fourier coefficients are related through convolutions of $\bar{\psi}(S)$.

\subsection{Choosing the computational basis}
\label{ChoosingTheComputationalBasis}
In the research of neural quantum states, the problem of how to choose the most suitable reference basis for a given quantum state remains unsolved. However, with the framework introduced above, we can intuitively draw conclusions about the desired NQS structure, which provides insights about potential difficulties for the quantum state representation.

The internal NQS structure can be analyzed by rotating the variational wave function in the correlator basis:
\begin{equation}
    \psi(\vec{\sigma}) = \sum_{S\subseteq [L]} \bar{\psi}(S) \mathcal{X}_S(\vec{\sigma}).
\end{equation}
Importantly, the coefficients $\bar{\psi}(S)$ depend solely on the variational parameters and not on the input configuration and are therefore a direct consequence of the network's parameters and structure (see Section~\ref{MathFramework}). To determine whether the network must learn a relatively simple or complex structure of coefficients $\bar{\psi}(S)$, it is necessary to identify known patterns or similarities.

To do so, we revisit Eq.~\eqref{Eq:Ham_corr}. In this equation we show the rotation of the Hamiltonian from the spin basis $|\sigma\rangle$ to the correlation basis $|\mathcal{X}_S\rangle$:
\begin{equation}
\hat{H} = \underbrace{\frac{1}{2^L}\sum_{\vec{\sigma},S} \langle\mathcal{X_S} |\vec{\sigma}\rangle |\mathcal{X}_S\rangle \langle \vec{\sigma}  |}_{\mathcal{X}} \hat{H} \underbrace{\frac{1}{2^L}\sum_{\vec{\sigma}',S'} \langle\vec{\sigma}'|\mathcal{X}_S' \rangle |\vec{\sigma}'\rangle\langle \mathcal{X}_S' | }_{\mathcal{X}^\dagger}.
\end{equation}
Although this rotation matrix $\mathcal{X}$ with matrix elements $\mathcal{X}_{S,\vec{\sigma}}=\langle\mathcal{X_S} |\vec{\sigma}\rangle $ can seem complicated, it has the simple and well known form of Hadamard matrices for spin-$1/2$ systems~\cite{ODonnell_2014}:
\begin{equation}
\begin{split}
    &\mathcal{X} = H_1 \otimes H_2 \otimes...\otimes H_L\\ &\mathrm{~with~} H_i = \left( \begin{matrix}
        \mathcal{X}_{S_i = \{\emptyset\},\uparrow} = 1& \mathcal{X}_{S_i = \{\emptyset\},\downarrow} = 1\\\mathcal{X}_{S_i = \{i\},\uparrow} = 1 &\mathcal{X}_{S_i = \{i\},\downarrow} =-1
    \end{matrix}\right)
\end{split}
\end{equation}
The same basis transformation is used to rotate from $|\vec{\sigma} = \vec{\sigma}_{z/x}\rangle $ to $|\vec{\sigma}_{x/z}\rangle$.
\begin{equation}
\begin{split}
    U = \sum_{\vec{\sigma}_x,\vec{\sigma}_z} u_{\vec{\sigma}_x,\vec{\sigma}_z} |\vec{\sigma}_x\rangle\langle \vec{\sigma}_z| = H^{\sigma}_1 \otimes H^{\sigma}_2 \otimes...\otimes H^{\sigma}_L\\
    \mathrm{~with~} H^{\sigma}_i = \left( \begin{matrix}
        u_{+,\uparrow} = 1& u_{+,\downarrow} = 1\\ u_{-,\uparrow} = 1&u_{-,\downarrow} =-1
    \end{matrix}\right)
\end{split}
\end{equation}
Consequently, the matrix elements $h^{\mathcal{X},z/x}_{n,m}$ of a Hamiltonian in the correlator basis $|\mathcal{X}_S\rangle$ are the same as the elements $h^{\sigma_{x/z}}_{n,m}$ in the $|\vec{\sigma}_{x/z}\rangle$ basis:
\begin{equation}
\begin{split}
        \hat{H} = &\sum_{n,m = 1}^L h^{\mathcal{X},z/x}_{n,m} |n_\mathcal{X} \rangle \langle m_\mathcal{X}| \\=& \sum_{n,m = 1}^L h^{\sigma_{x/z}}_{n,m} |n_{\sigma_{x/z}}\rangle \langle m_{\sigma_{x/z}}| ,
\end{split}
\end{equation}
with $h^{\mathcal{X},z/x}_{n,m} = h^{\sigma_{x/z}}_{n,m}$. Here, $n,m \leq 2^L$ are numbers that correspond to the $n^\mathrm{th}$ basis vector in the given basis. E.g. in the $|\sigma_z \rangle$ basis, the number $n =1$ ($n = 2^L$) corresponds to the basis state:
\begin{equation}
\begin{split}
    &|n_{\sigma_{z}}=1\rangle \equiv |\uparrow\uparrow...\uparrow\rangle = \bigotimes_{i=1}^{L}\left(\begin{matrix}
    1 \times|\uparrow_i\rangle\\0 \times|\downarrow_i\rangle
\end{matrix}\right)\mathrm{~and~}\\ &|n_{\sigma_{z}} = 2^L \rangle \equiv|\downarrow\downarrow...\downarrow\rangle  = \bigotimes_{i=1}^{L}\left(\begin{matrix}
    0\times|\uparrow_i \rangle\\1\times|\downarrow_i\rangle
\end{matrix}\right)
\end{split}
\end{equation}
and in the correlator basis $|\mathcal{X}_S\rangle$ we define it as:
\begin{equation}
\begin{split}
    &|n_\mathcal{X}  = 1\rangle \equiv |\mathcal{X}_\emptyset\rangle = \bigotimes_{i=1}^{L}\left(\begin{matrix}
    1 \times |\mathcal{X}_{S_i = \{\emptyset\}} \rangle\\0\times|\mathcal{X}_{S_i = \{i\}} \rangle
\end{matrix}\right)\mathrm{~and~}\\
&|n_\mathcal{X} =2^L\rangle\equiv|\mathcal{X}_L\rangle  = \bigotimes_{i=1}^{L}\left(\begin{matrix}
    0 \times |\mathcal{X}_{S_i = \{\emptyset\}}\rangle  \\1\times|\mathcal{X}_{S_i = \{i\}} \rangle
\end{matrix}\right).
\end{split}
\end{equation}
Here, we want to emphasize that the sites included in $S$ correspond to minus spins in the $\sigma^x$ basis:
\begin{equation}
\label{EQ:relationcorr_xbasis}
    |\mathcal{X}_S\rangle = \sum_{\vec{\sigma}^z} \prod_{i\in S} \sigma_i |\vec{\sigma}^z\rangle = \bigotimes_{i \notin S} \underbrace{\sum_{\sigma^z_i}|\sigma^z_i\rangle}_{|+_j\rangle} \bigotimes_{j \in S} \underbrace{\left(\sum_{\sigma^z_j} \sigma_j|\sigma^z_j\rangle \right)}_{|-_j\rangle}
\end{equation}
Note that we purposely added the index $z/x$ to clarify that the matrix element $h^{\mathcal{X},z/x}_{n,m}$ contains products of single spins in the $z/x$ basis and therefore dependent on the computational basis ($\sigma_x$ or $\sigma_z$), while the corresponding elements $h^{\sigma_{x/z}}_{n,m}$ are the usual matrix elements in the respective $x/z$ basis.

The consequence of having the same matrix elements $h^{\mathcal{X},{z/x}}_{n,m} = h^{\sigma_{x/z}}_{n,m}$ is that the coefficients for the ground state wave function are the same in correlator basis $\psi^{\mathcal{X},{z/x}}(n)$ and the corresponding spin basis $\psi^{\sigma_{x/z}}(n)$:
\begin{equation}
    \psi^{\mathcal{X}^{z/x}}(n)= \psi^{\sigma_{x/z}}(n),
\end{equation}
with:
\begin{equation}
    |\Psi\rangle = \sum_{n=1}^L \psi^{\mathcal{X},{z/x}}(n) |n_\mathcal{X}\rangle  = \sum_{n=1}^L \psi^{\sigma_{x/z}}(n) |n_{\sigma_{(x/z)}}\rangle,
\end{equation}
which allows us to predict the difficulty for NQS simulations. Note that we use $\psi^{\mathcal{X}^{z/x}}(n)$ instead of $\bar{\psi}(S)$  at the $n^\mathrm{th}$ basis vector of $|\mathcal{X}_S\rangle$ for better comparability.

\textit{What does this mean and how can we profit from this?---} 
In almost all cases, we intuitively pick the basis in which the wave function is peaked the most. E.g. the natural choice for an $|\uparrow \uparrow...\uparrow \rangle$ state would be the $|\sigma_z\rangle$ basis, since this provides an easy understanding of what we want to represent. However, as we showed in Section~\ref{Explaining_observations}, this requires surprisingly much computational effort (all correlation orders require the same coefficient), since it is difficult to align the weights such that most of the wave function coefficients are tuned to zero.

The analogy of correlations in the $|\vec{\sigma}_{z}\rangle$ basis to the wave function in the $|\sigma_x\rangle$ basis would have shown the same result. The product state $|\uparrow \uparrow...\uparrow \rangle$ in the $|\vec{\sigma}_{x}\rangle$ basis requires all wave function coefficients to be the same: $\psi^{\sigma_{x}}(n) = \psi^{\mathcal{X},{z}}(n) = C$. While the wave function in the $|\vec{\sigma}_{z}\rangle$ basis is peaked, the wave function in the correlator basis $|\mathcal{X}_S\rangle$ is widely spread.

If one uses the $|\vec{\sigma}_{x}\rangle$ basis instead, which is counterintuitive at first glance, the wave function in the correlator basis $|\mathcal{X}_S\rangle$ matches the wave function in the $|\vec{\sigma}_{z}\rangle$ basis and is therefore strongly peaked: $\psi^{\sigma_{z}}(n) = \psi^{\mathcal{X},{x}}(n) = \delta_{1,n}$. In a neural network, this requires just one single bias, with all other weights set to zero.

In summary, even a rough understanding of the ground state wave function in the $|\vec{\sigma}_{z}\rangle$ and $|\vec{\sigma}_{x}\rangle$ bases can provide a useful indication of which basis may lead to a wave function that is more or less peaked. Counterintuitively, choosing the reference basis where the true ground state wave function is less peaked will yield a more peaked wave function in the corresponding correlator basis. 
This can prove advantageous, especially when predominantly low correlation orders need to be captured by the NQS (compare with Section~\ref{Sec:CompTaylor}). However, one must keep in mind that the sign structure also changes with the basis, which can complicate the optimization~\cite{Bukov_2021,Schurov_2025}.

Note that by applying unitary transformations, often in the form of Pauli operators $U_{z/x} = \prod_{i\in{A}} \sigma_i^{z/x}$, we can change the sign structure $\psi_\mathrm{NQS}(\sigma) U_z |\sigma_z\rangle \rightarrow (-1)^{N^\sigma_A}\psi_\mathrm{NQS}(\sigma) |\sigma\rangle$ ($N^\sigma_A$ is the number of down spins on the sublattice A) and the order of the basis states $\psi_\mathrm{NQS}(\sigma) U_x|\sigma_z\rangle \rightarrow \psi_\mathrm{NQS} (\sigma) |-\sigma^{A}_z\rangle \otimes|\sigma^{\bar{A}}_z\rangle$. In many cases, this can help to construct a basis that reduces the requirements of the NQS wave function.

\textit{Remark.---} Non-local basis transformations, such as a transformation to the Gaussian basis~\cite{liu_2026} or the momentum basis~\cite{moreno_2023,Sobral_2025} aim to trade a decrease in complexity of representing a quantum state with increased computational costs.

\subsection{Entanglement bound from the expansion order}
\label{Consequences:Entanglement}
A typical measure for complexity of a quantum state is the entanglement entropy $S(\rho_A)$ of a subregion $A$. Based on the entanglement entropy, one usually defines three complexity scalings for quantum states: States with low entanglement entropy scaling (constant or logarithmic), states whose entanglement entropy scales with the area of the subsystem, and state with volume law scaling. In the following we provide an upper bound for the maximal entanglement entropy an NQS can capture, based on its maximal correlation order $n$.

\begin{figure}[t]
    \centering
    \includegraphics{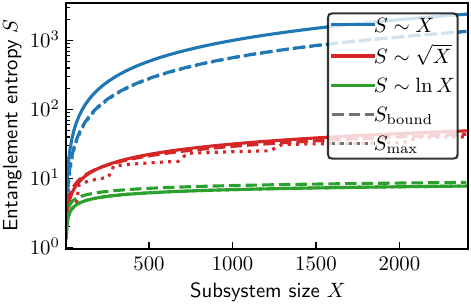}
    \caption{Entanglement scaling for different neural networks with maximal correlation order scaling with subsystem size $n = \frac{1}{4}X$ (blue), with the area $n = \sqrt{X}/\ln(X)$ (red), and kept constant $n = 1$ (green).}
    \label{Fig:Entanglement_entropy}
\end{figure}
The entanglement entropy $S(\rho_A)$ of a subregion $A$ is defined as:
\begin{equation}
    S(\rho_A) = -\mathrm{Tr}(\rho_A\ln(\rho_A)),
\end{equation}
with $\rho_A = \mathrm{Tr}_B|\psi\rangle \langle \psi| $. Rewriting $\rho_A$ in the correlator basis provides an intuitive insight into the complexity of representing a given quantum state:
\begin{equation}
\begin{split}
        (\rho_A)_{a,a'} & = \sum_B \psi(a,B) \psi^*(a',B)\\ & =  \sum_B \bar{\psi}(S_a,S_B) \bar{\psi}^*(S_{a'},S_B)\\
        & = (MM^\dagger)_{a,a'}.
\end{split}
\end{equation}
The rank of matrix $M$ determines the maximal entanglement entropy captured by a network with correlation order $n$:
\begin{equation}
\begin{split}
&S(n,\rho_A)\leq \ln(\mathrm{rank(M)}) \\  &\overset{X = \min(A,B)}{\leq}   S_\mathrm{max}(n,X)=\ln\left(\sum_{k = 0}^n {X\choose k}\right)\\ &\overset{n/X\leq 1/2}{\leq}S_\mathrm{bound}(n,X)= XH\left(\frac{n}{X}\right)
\end{split}
\end{equation}
where $H\left(\frac{n}{X}\right)$ is the binary entropy function~\cite{Cover_2005,galvin_2014}, valid for $n/X\leq 1/2$:
\begin{equation}
    S_\mathrm{bound}(n,X) 
    = - n \ln \left(\frac{n}{X} \right) - (X-n) \ln \left(1  - \frac{n}{X} \right).
\end{equation}
This upper bound reveals that the required minimal correlation order depends sensitively on the entanglement scaling of the desired quantum state. In the following, we analyze three distinct regimes, compare with Fig.~\ref{Fig:Entanglement_entropy}:
\begin{itemize}
    \item For a fixed expansion order $n$, the maximal entanglement entropy scales logarithmically $S \propto \ln X$.
    \item To represent states with area law entanglement $S \propto \sqrt{X}$, the correlation order must scale at least as $n \propto \sqrt{X}/\ln (X)$.
    \item For volume law entangled states $S \propto X$, the correlation order must scale at least as $n \propto X$.
\end{itemize}
The derivation of those scalings is shown in the Appendix~\ref{App:Entanglementscalings}. Consequently, representing generic many-body states requires a correlation order that increases with system size, with only weakly entangled states remaining accessible at fixed expansion order.

\textit{Remark.---} A large correlation order does not imply entanglement but can be understood as an entanglement capacity. However, the width and complexity of the respective spectrum of coefficients is related to entanglement.

\subsection{Relation to Taylor expansion}
\label{Sec:CompTaylor}
\begin{figure*}[t]
  \centering
    \includegraphics{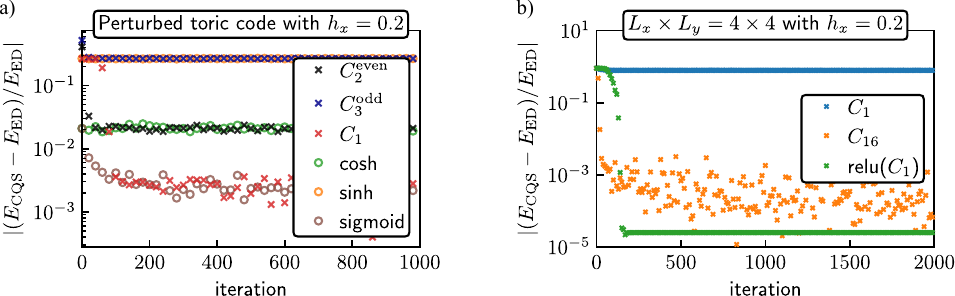}
\caption{
Convergence of NQS models applied to (a) the perturbed toric code (Gauss's law enforced in the sampling) and (b) the transverse field Ising model. In (a) we show a comparison between several CQS and a single layer FFNN with different activation functions. The CQS resemble low-order expansions of the corresponding activation functions ($\cosh \approx C_2^\mathrm{even}$, $\sinh \approx C_3^\mathrm{odd}$, $\mathrm{sigmoid} \approx C_1$). In (b), we compare the CQS ($C_1$ and $C_{16}$) to a 
single layer FFNN (equivalent to $C_1$) with rectified linear unit activation function.
}
\label{Fig:Fig_TaylorAndRelu}
\end{figure*}
In neural quantum states, a machine learning algorithm is trained such that it can represent a quantum state. In the previous sections, we addressed the topic of representability based on the correlations that the CQS has access to and compared it to the exact polynomial description of the desired quantum state given by its Fourier expansion. In the following, we want to generalize this to common neural network architectures, by using the complementary polynomial Taylor expansion of a neural network, and apply the information, we derived in the previous sections. Here, we distinguish between globally analytic NQS and NQS with only locally analytic functions before discussing general activation functions and the role of the bias.

\textit{Connection between Fourier and Taylor expansion: Globally analytic NQS.---}
The Taylor expansion of a function $\psi(\vec{\sigma})$ around point $\vec{p}$ is defined by:
\begin{equation}
    \psi(\vec{\sigma}) = \sum_{n=0}^\infty \frac{\psi^{(n)}(\vec{p})}{n!} (\vec{\sigma}-\vec{p})^n.
\end{equation}
Assuming that we work with globally analytic functions, the full NQS can be expanded around $\vec{p} = \vec{0}$ up to some significant order $k$:
\begin{equation}
\begin{split}
    \psi (\vec{\sigma}) \approx  ~& \psi(\vec{0}) + \sum_i \left. \frac{\partial \psi (\vec{p})}{\partial p_i} \right|_{\vec{p}=0} \sigma_i + ...\\ &  + \frac{1}{k!}\sum_{i_1,...,i_k}\left. \frac{\partial^k \psi (\vec{p})}{\partial p_{i_1} ...\partial p_{i_k}} \right|_{\vec{p}=0} \sigma_{i_1}...\sigma_{i_k},
\end{split}
\end{equation}
where the relevance of the remainder $R_k(\vec{\sigma})=\sum_{n=k+1}^\infty \left.\frac{\psi^{(n)}(\vec{p})}{n!} \right|_{\vec{p}=0}\vec{\sigma}^n$ is suppressed by the factorial. As both, Taylor and Fourier expansion use the same polynomial structure, we can argue that the highest relevant Taylor expansion order defines the highest accessible correlation order. 

In the previous sections we showed that high correlation orders remain relevant: a) peaked wave functions require broad support of the correlation weights, such that even the highest correlation order $\bar{\psi}(L)$ can be a significant contribution: $\bar{\psi}(L) \geq \bar{\psi}(S)$ (Section~\ref{Explaining_observations}) and b) depending on the entanglement scaling of the desired state, there exists a minimal relevant expansion order that depends, for most quantum states, on the system size (Section~\ref{Consequences:Entanglement}). 

Consequently, NQS-like quantum state representation requires high, non-vanishing Taylor expansion orders (often of order system size) to be able to represent a quantum state faithfully. In practice, this mismatch between factorial suppression and requiring high expansion orders necessitates:
\begin{itemize}
    \item the network to increase the magnitude of the variational parameters to amplify the contributions from higher-order correlations,
    \item the use of deeper neural networks that are more efficient at generating high-order correlations with small weights.
\end{itemize}

We want to highlight that this conclusion immediately leads to a significant distinction between increasing the width and increasing the depth of a neural network: While wider neural networks still have to rely on increasing the magnitude of the weights to generate higher correlation orders, deeper neural networks achieve this naturally by a consecutive use of activation functions.

Furthermore, we emphasize that requiring large network weights potentially comes with a drawback: stochastic reconfiguration, which is based on linearized imaginary time evolution, is designed for small steps in both the imaginary time evolution and the parameter manifold. Therefore, large weights can slow down the accurate optimization.

As an example, consider representing an $|\uparrow \uparrow...\uparrow \rangle$ state ($|\sigma_z\rangle$ basis) with a simple single layer feed forward neural network and a $\cosh$ activation function. An exact representation of this simple product state would require infinitely large network parameters $W_i$:
\begin{equation}
    \psi(\vec{\sigma}) = \cosh(W_1\sigma_1 + W_2\sigma_2+ ... +W_L\sigma_L ).
\end{equation}
A more intuitive way to think about this is that the network needs to maximize the distance between $\psi(\uparrow \uparrow...\uparrow )$ and all other wave function coefficients, which can only be achieved in the large $W_i$ limit.

In combination with Section \ref{ChoosingTheComputationalBasis}, the Taylor expansion perspective also suggests an intuitive picture of how common NQS initializations explore the Hilbert space. For architectures initialized with small weights, the initial state is often close to the uniform superposition $|++...+\rangle$ state, as their Taylor expansion is dominated by low order correlations. Small step optimization techniques, like stochastic reconfiguration, then gradually deform the wave function in the Hilbert space. During the optimization higher-order terms in the Taylor expansion become increasingly relevant. Through the Hadamard relation, this process gets a second equivalent interpretation in the $\sigma^x$ basis: as shown in Eq. \eqref{EQ:relationcorr_xbasis} the correlation order $|S|$ corresponds to a set of $\sigma^x$ basis states with $|S|$ minus spins. Therefore, the hierarchy of the correlation order given by the Taylor expansion implies a bias that NQS naturally explore basis states with few minus spins first and potentially increase the number of minus spins during the training.\\

\textit{Locally analytic NQS.---}
Next, we comment on NQS with activation functions that are only locally analytic. These functions have a finite convergence radius and, thus, do not necessarily have the same polynomial behavior for all basis states. However, this can simplify the optimization. Locally analytic functions have at least two regimes with distinct behavior. This allows us to separate or group basis states by shifting them in different regimes. A prominent example for locally analytic functions is the rectified linear unit (ReLu):
\begin{equation}
    \mathrm{ReLu}(x) = \left\{ \begin{matrix}
        x & \mathrm{for~} x>0,\\
        0 & \mathrm{for~} x\leq0. \\ 
    \end{matrix}
    \right.
\end{equation}
As can be readily seen, the ReLu will not increase the order of correlations in the system; nonetheless, it is a very efficient activation function. Instead of building up correlations, it has the ability to group states by tuning the bias, which allows us to set probabilities of multiple basis states to zero. The system can then use all the other network parameters to represent the wave function coefficients of the remaining states. An example for a transverse field Ising model ($\hat{H}_\mathrm{TFIM}  = \hat{H}_\mathrm{Ising} -h_x\sum_i \hat{\sigma}^x_i $) on a $L \times L = 4 \times 4 $ lattice is shown in Fig.~\ref{Fig:Fig_TaylorAndRelu}\,b). Here, the feed forward neural network achieves good results by tuning the bias such that the ReLu activation sets the wave function coefficients of most basis states to zero.  Specifically, the wave function has nonzero coefficients for only 17 basis states (out of $\mathcal{O}(10^5) $). This is not a coincidence, since the NQS has access to precisely 17 correlations (one bias ($\mathcal{X}_\emptyset = 0^\mathrm{th}$ order) and $L\times L = 16$ first order correlations, compare with Eq.~\eqref{Eq:Num_Corrs}). In the context of a system of equations, introduced in Section~\ref{Sec_intuitiveFE}, this implies that, when learning the perturbed AFM state, the network can faithfully solve the set of equations for at most 17 basis states. Similarly, other locally analytic functions like $\tanh$ or $\mathrm{sigmoid}$ have the ability to group basis states, such that the optimization can potentially be simplified.\\

\textit{Even and odd orders: The role of the bias.---}
As shown in Section~\ref{MathFramework}, any wave function of a finite size spin system can be described as a combination of even and odd orders of correlations: $\psi(\sigma) = Y_\mathrm{even}(\sigma)  + Y_\mathrm{odd}(\sigma)$. While even orders are spin symmetric ($Y_\mathrm{even}(\sigma) = Y_\mathrm{even}(-\sigma)$), odd orders are antisymmetric ($Y_\mathrm{odd}(\sigma) = -Y_\mathrm{odd}(-\sigma)$). Therefore, the quantum states with an even (odd) parity (i.e. a ground state of the Ising model), can be represented by solely using even (odd) correlations. On the other hand, systems that are not spin symmetric $|\psi(\sigma)| \neq |\psi(-\sigma)|$ require both even and odd orders.

While this is not surprising, one has to keep in mind that many frequently used activation functions do have a fixed parity:
\begin{align}
\begin{split}
    \sinh(x) &= -\sinh(-x),\\
    \cosh(x) &= +\cosh(-x),\\
    \tanh(x) &= -\tanh(-x),\\
    \arctan(x) &= -\arctan(-x),\\
        &~\,\vdots\\
    \frac{1}{2}- \mathrm{sigmoid}(x) &= -(\frac{1}{2}-\mathrm{sigmoid}(-x)).\\
\end{split}
\end{align}
Therefore, network architectures that maintain the parity of these activation functions cannot represent quantum states with a different parity. 
Note that the Taylor expansion of functions with even (odd) parity exclusively contains correlations of even (odd) order. To illustrate this we apply different network architectures to the in Section~\ref{Observations} introduced perturbed toric code model, where Gauss's law is enforced in the sampling. In Fig.~\ref{Fig:Fig_TaylorAndRelu}\,a) we show the convergence of a single layer feed forward NQS with different activation functions ($\cosh$, $\sinh$, and $\mathrm{sigmoid}$) and compare them to the correlator transformer with access to either purely even ($C_2^\mathrm{even}$), odd ($C_3^\mathrm{odd}$) or $ 0^\mathrm{th} + \mathrm{~odd} $ order ($C_1$) correlations. Although all NQS do not fully capture the ground state, the behavior of the activation functions can be understood very well by comparing them to the CQS, as those resemble their low expansion orders ($\cosh \approx C_2^\mathrm{even}$, $\sinh \approx C_3^\mathrm{odd}$, $\mathrm{sigmoid} \approx C_1$). Since increasing the correlation order of just one parity does not improve the result (the activation functions contain all orders of one parity), we can attribute the inability to converge to the absence of interplay between even and odd correlations.

Although this may appear to restrict the use of certain activation functions in most models, these limitations can be easily overcome by introducing a bias in the neural network layer. The bias enables the presence of odd (even) correlation orders in an otherwise even (odd) order of the Taylor expansion, making it not merely a convenient optimization parameter but a crucial component for an accurate quantum state representation.

\section{Comment on Network architectures} 
\label{Comment_on_architectures}
In the following, we want to use the acquired knowledge about the working principle of NQS and their representational power to comment on potential advantages and disadvantages of commonly used network architectures.\\

\textit{Autoregressive NQS.---}
In the research of neural quantum states, two directions have emerged, namely non-autoregressive methods based on Metropolis Hastings Monte Carlo, where the machine learning model outputs the full wave function at once, and autoregressive NQS that evaluate the amplitude of the full wave function as a product of local, conditional probabilities:
\begin{equation}
\label{Eq:auto_def}
    p(\vec{\sigma}) = p(\sigma_1)p(\sigma_2|\sigma_{\leq1})...p(\sigma_L|\sigma_{\leq L-1}).
\end{equation}
Thus, the machine learning model gradually acquires information about the input and has to adjust its conditional probabilities accordingly. Each conditional probability $p(\sigma_k|\sigma_{\leq k-1})$ can be written in its orthogonal correlator basis according to Eq.~\eqref{Eq:WaveFunc}:
\begin{equation}
\label{Eq:auto_corr_basis}
    p(\sigma_k|\sigma_{\leq k-1}) = \sum_{S \subseteq [k]} \bar{p}_k(S) \mathcal{X}_S(\sigma),
\end{equation}
where $\bar{p}_k(S)$ is defined as:
\begin{equation}
\label{Eq:auto_cond}
    \bar{p}_k(S) = \frac{1}{2^k}  \sum_{\vec{\sigma}\in \{-1,1\}^k} p(\sigma_k|\sigma_{\leq k-1}) \mathcal{X}_S(\sigma).
\end{equation}
Therefore, the total probability $p(\vec{\sigma})$ is by design a construction that contains all correlation functions, which is required for a good approximation of many quantum states. Thus, the relevant criterion for autoregressive NQS is the difficulty to represent the conditional probabilities $p(\sigma_k|\sigma_{\leq k-1})$, which has a worst case complexity of $\mathcal{O}(2^k)$ (in total $\mathcal{O}(2^{\sum_k^L k})$). However, many quantum state satisfy conditional independence:
\begin{equation}
\label{Eq:auto_indep}
    p(\sigma_k|\sigma_{k-1\geq }) = p(\sigma_k|\sigma_{k-1},\sigma_{k-2},...,\sigma_{k-l}),
\end{equation}
with small $l$, which reduces the complexity of the full problem to approximately $\mathcal{O}(L\times2^l)$, such that an efficient representation can be found~\cite{yang_2024}. The term short ranged conditional probabilities can be translated to our formalism in Eq.~\eqref{Eq:auto_cond}, as it is equivalent to setting the correlation order coefficient $\bar{p}_k(S)$ to zero, if $S$ contains any $\sigma_j$ with $j\leq k-l-1$.

Note that autoregressive models have an inherent asymmetry in the architecture due to their path-like structure (see Eq.~\eqref{Eq:auto_def})~\cite{Doeschl_2025}, as the first site only depends on itself, while the probability on the last site depends on all sites. This will also become obvious when writing Eq.~\eqref{Eq:auto_def} in terms of $\bar{p}_k(S)$ and $\mathcal{X}_S(\sigma)$. The more short ranged a conditional probability is (compare with Eq.~\eqref{Eq:auto_indep}), the more symmetric the equation becomes. Furthermore, the path-like structure predefines the required decision tree in truncated Hilbert spaces, allowing for only one possible polynomial solution (compare with Section~\ref{DecisionTreeForm}).\\

\textit{Non-autoregressive NQS.---}
Non-autoregressive NQS model the full wave function directly, without factorizing it into a sequence of conditional probabilities. Therefore, this requires network architectures to be capable of generating high correlations orders. This is reflected in the development of the current research direction. Early approaches often relied on Restricted Boltzmann machines, due to their analytical accessibility~\cite{Gao_2017,Carleo_2018}. However, the lack of complexity results in an inefficient way to generate large coefficients for high correlation orders. This limitation has unwittingly led to the adoption of architectures that can naturally generate high correlation orders with large coefficients.

Transformer neural quantum states are a prominent example~\cite{Wu_2023,Rende_2024_LinAl,Viteritti_2023,Viteritti_2025,Lange_2025}: the multiplicative terms in the attention mechanism naturally produce large coefficients for higher correlation orders:
\begin{equation}
    \mathrm{Attention}(Q ,K,V) = \mathrm{softmax}\left(\frac{QK^T}{\sqrt{d_k}}\right)V,
\end{equation}
where $Q,K,V$ are all linear in the input, which itself can be a sum over multiple correlation orders.

Similarly, determinant-based architectures have become the unchallenged architecture for fermionic~\cite{Moreno_2022,Chen_2025,Wurst_2025} systems. 
To illustrate that determinant wave functions generate large coefficients for high correlation orders efficiently, we discuss neural backflow based determinant constructions for $m$ fermions on $L$ sites~\cite{Lu_2019,Becca_Sorella_2017}:
\begin{equation}
\begin{split}
    \Psi(\vec{n}) =& \det \left| \begin{matrix}
        \phi^\mathrm{bf}_{k_11}(n_1,...n_L)&...&\phi^\mathrm{bf}_{k_1m}(n_1,...n_L)\\
        \vdots&\ddots &\vdots\\
        \phi^\mathrm{bf}_{k_m1}(n_1,...n_n)&...&\phi^\mathrm{bf}_{k_mm}(n_1,...n_L)
    \end{matrix} \right|,
\end{split}
\end{equation}
where $\phi^\mathrm{bf}_{k_ij}(n_1,...n_n)$ are backflow enhanced elements. $k_i$ denotes the index of the $i^\mathrm{th}$ occupied site and $j$ represents the $j^\mathrm{th}$ fermion. In a backflow based determinant, one introduces the potentially deep neural network $f^\theta_{k_ij}(n_1,...n_n)$ that uses Fock space configurations as inputs, at an orbital level:
\begin{equation}
    \phi^\mathrm{bf}_{k_ij}(n_1,...n_n) = \phi_{k_ij} + f^\theta_{k_ij}(n_1,...n_n).
\end{equation}
Therefore, $\phi^\mathrm{bf}_{k_ij}(n_1,...n_n)$ itself, can be a function with higher-order correlations. Importantly, however, the determinant structure acts as a further non-linear activation function: by involving products of system size many orbital, it automatically generates large coefficients for higher-order correlations:
\begin{equation}
    \Psi(\vec{n}) = \sum_{i_1,...,i_m} \epsilon_{i_1,...,i_m} \phi^\mathrm{bf}_{k_1,i_1}...\phi^\mathrm{bf}_{k_m,i_m}.
\end{equation}
A natural consequence can be that the specific choice of neural network architecture is less critical~\cite{lange2024simulatingtwodimensionaltjmodel}.

Since both network architectures are designed to naturally generate large coefficients for high correlation orders, they appear— from the correlator’s perspective— to be highly suitable for NQS simulations. This has been confirmed in numerous numerical benchmarks, where they outperformed other architectures~\cite{Luo_2019,chen_2025_CNN,Rende_2024,Choo_2020,Zakari_2025}.

\section{Conclusion}
In this work, we explored the importance of correlations in neural quantum states (NQS), in particular, how they determine a network's ability to faithfully represent quantum states. By using correlation based neural networks, we were able to numerically demonstrate that even simple quantum states, such as product states, can require all possible correlation orders for an accurate representation, whereas other, more complex quantum states can be represented using a simple correlation structure in a restricted Hilbert space.

Using tools from Boolean function theory, we provided a framework to better understand the requirements for NQS-like quantum state representation. This framework reveals that the internal NQS structure effectively represents a quantum state in the correlator basis and, consequently, requires a basis transformation from the computational basis, which can lead to non-trivial correlation order structures.

Furthermore, we observed a significant relaxation of correlation requirements in the case of truncated Hilbert spaces. Restricted Hilbert spaces result in linear dependencies, which allow CQS to store information about high correlation orders in significantly lower correlation orders. Additionally, we showed connections between spin basis rotations and the correlator basis, which can provide valuable insights into the desired NQS structure and potential problems for quantum state representation.

By using Boolean function theory, we derived an entanglement bound for a given maximal correlation order, which reveals that a fixed expansion order produces logarithmic entanglement scaling, while quantum states with area law or volume law entanglement require a quasi area or volume expansion order scaling, respectively. This has direct consequences on the neural network, as the highest relevant Taylor expansion order defines the highest accessible correlation order. To achieve these large expansion orders, neural networks need to have large variational parameters unless they are deep enough such that a consecutive use of activation functions generates high-orders naturally.

Finally, we investigated how activation functions and neural network architectures influence the expressiveness of the neural quantum state ansatz. These design choices affect the ability to group certain basis states or to generate large coefficients for high-order terms, which strongly influences the correlations that can be captured, and consequently, the network's representational power.

Overall, our results offer new insights into the internal structure of NQS and allow a deeper understanding of their requirements. By identifying the role of correlations in neural quantum states and clarifying how these correlations emerge from the network architecture, we provide a foundation for a systematic tailoring and optimization of NQS to a desired quantum system.

\textit{Note added.---} While finishing the manuscript, we became aware of a related work by Schurov et al.~\cite{Schurov_2025}, studying the complexity of sign structures using Boolean Fourier analysis.

\begin{acknowledgments}
We thank F. Pauw, T. Vovk, C. Roth, and D. Aliverti-Piuri for fruitful discussions. This research was supported by LMUexcellent, funded by the Federal Ministry of Education and Research (BMBF) and the Free State of Bavaria under the Excellence Strategy of the Federal Government and the Länder and by the Deutsche Forschungsgemeinschaft (DFG, German Research Foundation) under Germany’s Excellence Strategy (EXC-2111 -- 390814868).
\end{acknowledgments}

\section*{Code and data availability}
The code and data used to produce the results presented in this paper are available in the following GitHub repository: \url{https://github.com/FabianDoeschl/CoTra_NQS}

\clearpage
\appendix
\section{Correlator transformer: Detailed description}
\label{App:CoTra}
A major disadvantage of neural quantum states (NQS) is the limited understanding of their optimization and representation processes. Although we have techniques that minimize the energy or evolve the state in time (and so on), we have no intuition why certain parameters might be relevant or irrelevant.

To investigate this behavior, we employ a correlator transformer as an architecture for neural quantum states. The correlator transformer was introduced in Suresh et al.~\cite{Suresh_2024}. The model uses standard patched input configurations $\sigma_{i,p}$ with a linear embedding:
\begin{equation}
X^1_{i,m} = \sum_p\sigma_{i,\mathrm{p}}\,W_{\mathrm{p},m}\,\mathrm{pE}_{i,m}, \qquad X^1 \in \mathbb{R}^{(L_x/p,L_y/p,2p)},
\end{equation} 
where $i$ indexes the patch, $p$ labels the position within a patch, and $m$ denotes the embedding dimension. Here, $W_{\mathrm{p},m}$ are trainable network parameters and $\mathrm{pE}_{i,m}$ represents the standard positional encoding:
\begin{equation}
\mathrm{pE}_{i,m} = 
\left\{   
\renewcommand\arraystretch{1.2}
\begin{matrix}
\sin\left( \frac{i}{10000^{2j/d}}\right) & m = 2 j, \\
\cos\left( \frac{i}{10000^{2j/d}}\right) & m = 2j+1.
\end{matrix}
\right.
\end{equation}
The matrix $X^1_{i,m}$ that contains first order correlations is then used to calculate the query $Q_n$ and key $K_n$ values of the transformer:
\begin{equation}
    Q_n = X^1W^{nQ} \quad \mathrm{and} \quad K_n = X^n W^{nK},
\end{equation}
the value matrix, on the other hand, is given by: $V_n = \mathrm{pE} W^{nV}$
where $W^{nQ}$, $W^{nK}$ and $W^{nV}$ are network weights. A product of query, key, and value matrices defines the output of the corresponding layer: 
\begin{equation}
X^{n+1} = \frac{1}{\sqrt{2p}} Q_nK_n^TV_n.
\end{equation} 
All outputs $\mathbb{X} = [\bar{X}^1,\bar{X}^2,\dots,\bar{X}^i ,\dots,\bar{X}^n ]$, with $\bar{X}^i = \mathrm{mean}(X^i)$ are then used in a linear layer, with weights $W^{\mathrm{pred}}$, to predict the probability of the corresponding input configuration:
\begin{equation}
    \psi(\vec{\sigma}) = W^{\mathrm{pred}}\mathbb{X}(\vec{\sigma}) +\beta.
\end{equation} 
This wave function of polynomial form allows us to interpret the processes within the transformer. Here, we included the bias $\beta$ as this corresponds to $0^\mathrm{th}$ order correlations.

\textit{Number of parameters.---} 
To ensure a fair comparison between different correlation orders, we keep the total number of trainable parameters approximately constant. The total number of parameters depends on the system size ($N_x \times N_y$), the patch size $ p_{\mathrm{size}}$, and the embedding dimension $e_\mathrm{dim}$. 

The equation that defines the number of parameters is given by:
\begin{align}
\begin{split}
    \#\mathrm{params} = (\underbrace{p_{\mathrm{size}} e_\mathrm{dim}}_\mathrm{embedding} +\underbrace{\frac{N_xN_y}{p_{\mathrm{size}}}e_\mathrm{dim}}_\mathrm{\mathrm{pE}}) \\+  ~\underbrace{3 (n_\mathrm{lay} -1)e_\mathrm{dim}^2}_\mathrm{decoder} + \underbrace{n_\mathrm{lay} + 1}_{W^{\mathrm{pred}} +\beta}
\end{split}
\end{align}

\textit{Optimization.---} To optimize the NQS, we use standard stochastic reconfiguration techniques~\cite{Lange_2024} (with a learning rate of $\alpha = 0.01$) provided by the NetKet package~\cite{netket2:2019}.

\section{Generalizing to other encodings and non-spin systems}
\label{App:Generalization}
In the main text we focused our analysis on spin-$1/2$ systems. However, the idea to expand systems in terms of monomials of the input can be extended to systems with one-hot encoding or to cases with higher local Hilbert space dimensions.\\

\textit{One-hot encoding.---} 
We start by discussing one-hot encoding for general systems. This method embeds the Hilbert space of the system in a much larger one and truncates it afterward. To provide an intuitive example, we stick to a system with $\{-1,1\}^L$ spins. When using one-hot encoding, the effective Hilbert space is given by: $\{\{0,1\},\{0,1\} \}^L$. Since we are not interested in the whole effective Hilbert space, we truncate it and use the decision tree formalism. The indicator polynomial of the extended local Hilbert space is given by:
\begin{align}
\begin{split}
    &\mathbb{1}_{\alpha,\beta}(\sigma^s) = \mathbb{1}_{\alpha}(\sigma^s_1)\mathbb{1}_{\beta}(\sigma^s_2)  =\\ &\sum_{\alpha,\beta \in\{0,1\}^L} \frac{1-4(\alpha-\frac{1}{2})(\sigma^s_1-\frac{1}{2})}{2}\frac{1-4(\beta-\frac{1}{2})(\sigma^s_2-\frac{1}{2})}{2},
\end{split}
\end{align}
which reduces to:
\begin{equation}
    \mathbb{1}_{\alpha/\beta}(\sigma^s_{1/2}) = \left\{ \begin{matrix}
        \sigma^s_{1/2} & \mathrm{for~} \alpha/\beta = 1,\\
        1-\sigma^s_{1/2} & \mathrm{for~} \alpha /\beta= 0.
    \end{matrix}
    \right.
\end{equation}
This produces a polynomial similar to Eq.~(12) (considering that $\sigma^s_1 = 1-\sigma^s_2$). Hence, one hot encoding does not offer an immediate advantage (or disadvantage) compared to typical spin encoding. Note that the monomials are no longer orthogonal. Nevertheless, this expression can be useful, as it allows a relatively simple understanding of how it groups states: Terms of degree $k$ contribute to the coefficient of $2^{L-k}$ basis states. Therefore, characterizing a basis state requires one system size order correlation. E.g. a product state of up spins is given by $\psi_{\uparrow}(\vec{\sigma}) =\prod_{i=1}^L \sigma_{2,i}$, whereas down spins are given by $\psi_{\downarrow}(\vec{\sigma}) = \prod_{i=1}^L \sigma_{1,i}$.

\textit{Non binary local Hilbert space dimension.---} Finally, we turn to systems with larger local Hilbert spaces and non binary encoding. To do so, we introduce the Lagrange basis, a more general form of the indicator polynomial~\cite{Waring_1779,Bayen_2015,lindsey2024}. It defines an orthonormal basis $\{l_1(x^s),l_2(x^s),...,l_k(x^s)\}$ at site $s$ for a local Hilbert space dimension $\{x^s_1,x^s_2,...,x^s_k\}$ of $k$. The polynomial $l_{\alpha^s}(x^s)$ is defined by:
\begin{equation}
    l_{\alpha^s_j}(x^s) = \prod_{\underset{m\neq j}{0\leq m\leq k}} \frac{x^s-\alpha^s_m}{\alpha_j^s-\alpha^s_m} = \left\{ \begin{matrix} 1 & \mathrm{for~} x^s = \alpha^s,\\
    0 &\mathrm{else.}        
    \end{matrix} \right. 
\end{equation}
with $x^s,\alpha^s_j \in \{x^s_1,x^s_2,...,x^s_k\}$. The complete orthonormal basis for all sites $S$ is then defined by:
\begin{equation}
    L_{\vec{\alpha}}(\vec{x}) =\prod_{s \in S} l_{\alpha^s}(x^s),
\end{equation}
which allows us to write the wave function as:
\begin{equation}
    \psi(\vec{x}) = \sum_{\vec{\alpha} \in \{x_1,...,x_k\}^n} \psi(\vec{\alpha}) L_{\vec{\alpha}}(\vec{x}).
\end{equation}
Thus, the resulting wave function is a polynomial of degree $(k-1)\times n$. Although the analysis in this regime becomes more challenging, as the maximal polynomial order grows much faster with system size and its monomials are no longer orthogonal, we can still conclude that even for simple product states high expansion orders are required.

\section{Spanning tree formalism applied to the toric code model restricted to the physical sector}

\label{App:SpanningTree}

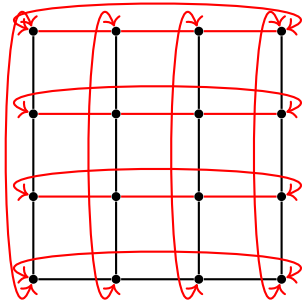
\begin{figure}
    \centering
\begin{tikzpicture}[scale=1.1, every node/.style={circle,fill,inner sep=1.2pt}]

\foreach \x in {0,...,3} {
  \foreach \y in {0,...,3} {
    \node (v\x\y) at (\x,\y) {};
  }
}

\draw[thick] (v00) -- (v10);
\draw[thick] (v10) -- (v20);
\draw[thick] (v20) -- (v30);

\draw[thick] (v00) -- (v01);
\draw[thick] (v01) -- (v02);
\draw[thick] (v02) -- (v03);

\draw[thick] (v10) -- (v11);
\draw[thick] (v11) -- (v12);
\draw[thick] (v12) -- (v13);

\draw[thick] (v20) -- (v21);
\draw[thick] (v21) -- (v22);
\draw[thick] (v22) -- (v23);

\draw[thick] (v30) -- (v31);
\draw[thick] (v31) -- (v32);
\draw[thick] (v32) -- (v33);

\draw[thick, red] (v01) -- (v11);
\draw[thick, red] (v11) -- (v21);
\draw[thick, red] (v21) -- (v31);

\draw[thick, red] (v02) -- (v12);
\draw[thick, red] (v12) -- (v22);
\draw[thick, red] (v22) -- (v32);

\draw[thick, red] (v03) -- (v13);
\draw[thick, red] (v13) -- (v23);
\draw[thick, red] (v23) -- (v33);

\foreach \y in {0,1,2,3} {
  \draw[<->,thick, red, bend right=160] (v3\y) to (v0\y);
}

\foreach \x in {0,1,2,3} {
  \draw[<->,thick, red, bend right=160] (v\x3) to (v\x0);
}

\end{tikzpicture}
    \caption{Spanning tree (black links) for a toric code model restricted to the physical sector. The $N_x \times N_y + 1$ spins on the red links span the full orthogonal basis, spins on black links are determined by the configuration of the red links due to Gauss's law.}
    \label{Fig:TC_Restricted_HS}
\end{figure}

Studying the perturbed toric code model is a difficult task as the Hilbert space grows with $2^{2(L_x\times L_y)}$. However, restricting the analysis to the physical sector, where Gauss's law is always fulfilled, reduces the effective Hilbert space dimension to $2^{L_x\times L_y+1}$. Following the spanning tree formalism allows us to limit the number of relevant spins to $L_x\times L_y+1$~\cite{Lang_2012,Sheffer_2025,zhao_2025}. This is shown in Fig.~\ref{Fig:TC_Restricted_HS}. Spins on the red links determine the full configuration and span the complete orthogonal basis (according to Eq.~(14)). 
Consequently, the remaining $L_x\times L_y-1$ spins on the black links are linearly dependent on the ``basis spins'' and thus contain relevant information about potentially higher-order correlations of the $L_x\times L_y+1$ basis spins.

Note that the spanning tree shown in Fig.~\ref{Fig:TC_Restricted_HS} is not unique: translations and rotations of this tree will yield an equally well-suited orthogonal basis. 

\section{Relation between $\psi(\sigma)$, $p(\sigma)$, and $\log \psi(\sigma)$}
\label{App:relationSpinmomentsandlogpsi}
In this section, we clarify the relationship between three representations of a quantum state that appear naturally in the literature: $\psi(\sigma)$, $p(\sigma) = |\psi|^2$, and $\log \psi(\sigma)$. Although related, these objects have generally different Fourier structures.
\subsection{Probability and spin-moments}
The probability distribution of a quantum many body state is given by the square of the wave function $p(\sigma)= |\psi(\sigma)|^2$ and therefore describe similar things. In particular, in some autoregressive neural quantum states, the network output directly parametrizes $p(\sigma)$, which raises the question how both are related from a boolean function point of view.
The Fourier coefficients $\bar{p}(S) $ of the probability $p(\sigma) = \sum_{S \subseteq [L]} \bar{p}(S) \mathcal{X}_S(\sigma)$ are given by:
\begin{equation}
\begin{split}
\bar{p}(S) &= \frac{1}{2^L} \sum_\sigma \psi^*(\sigma)\psi(\sigma) \chi_s(\sigma) \\ &=  \frac{1}{2^L} \sum_\sigma \sum_{S',S''}\bar{\psi}^*(S')\bar{\psi}(S'') \chi_s(\sigma) \chi_{s'}(\sigma) \chi_{s''}(\sigma).
\end{split}
\end{equation}
Using $\chi_A(\sigma)\chi_{A+B}(\sigma)= \chi_B(\sigma)$ and $\frac{1}{2^L} \sum_\sigma \chi_{A}(\sigma) \chi_{B}(\sigma)  =  \delta_{A,B}$, this equation reduces to:
\begin{equation}
    \begin{split}
\bar{p}(S) &=  \frac{1}{2^L} \sum_\sigma \sum_{S',\delta}\bar{\psi}^*(S')\bar{\psi}(S'\Delta \delta) \chi_s(\sigma) \chi_{\delta}(\sigma) \\&= \sum_{S' \subseteq  [L]}\bar{\psi}^*(S')\bar{\psi}(S\Delta S'),
\end{split}
\end{equation}
where $\Delta$ denotes the difference between sets. Consequently, the Fourier coefficients $\bar{p}(S)$ is defined as a convolution of the $(\bar{\psi}^* * \bar{\psi})(S) = \sum_{S' \subseteq  [L]}\bar{\psi}^*(S')\bar{\psi}(S\Delta S')$.
Thus, although there exists a clear relation between $p(\sigma)$ and $\psi(\sigma)$, their Fourier coefficients may differ significantly from one another.

\textit{Remark.---} The Fourier coefficients $\bar{p}(S)$ can be identified as spin moments $ \langle \prod_{i \in S} \hat{\sigma}^z_i \rangle =  \frac{1}{2^L} \sum_\sigma \psi^*(\sigma)\psi(\sigma) \chi_s(\sigma)$ and therefore have a direct physical interpretation.

\subsection{Logarithmic parametrization of the neural quantum states}
In practice, neural quantum states are often defined by using $\log\psi(\sigma) = f(\sigma)$ instead of $\psi(\sigma) = f(\sigma)$, where $f(\sigma)$ is the neural network implementation. This trick is often applied to gain numerical stability, as $\psi(\sigma)$ can vary over many orders of magnitude, which is related to requiring large correlation orders. Although, the logarithmic implementation can be understood as adding another activation function, the question arises, how their expansion coefficients $\bar{f}(S)$ and $\bar{\psi}(S)$ are related:
\begin{equation}
\begin{split}
    \bar{\psi}(S) =& \frac{1}{2^L} \sum_\sigma \exp(f(\sigma)) \mathcal{X}_S(\sigma)\\
    =& \frac{1}{2^L} \sum_\sigma \left(\sum_{n=0}^{\infty} \frac{1}{n!} f(\sigma)^n\right)  \mathcal{X}_S(\sigma)\\
    =& \sum_{n=0}^{\infty} \frac{1}{n!} \sum_{S_1,...,S_n\subseteq[L]}\bar{f}(S_1) ...\bar{f}(S_n)\\ &\underbrace{\frac{1}{2^L} \sum_\sigma \mathcal{X}_{S_1}(\sigma)...\mathcal{X}_{S_n}(\sigma)\mathcal{X}_{S}(\sigma)}_{\delta_{S_n, S\Delta S_1\Delta...\Delta S_{n-1}}}\\
    =&\sum_{n=0}^{\infty} \frac{1}{n!} \sum_{S_1,...,S_{n-1}\subseteq[L]}\bar{f}(S_1) ...\\ &\bar{f}(S_{n-1})\bar{f}(S\Delta S_1\Delta...\Delta S_{n-1}) \\
    = &\sum_{n=0}^{\infty} \frac{1}{n!} \bar{f}^{*n}(S).
    \end{split}
\end{equation}
Therefore, $\bar{\psi}(S)$ is a sum of all possible $n$-fold convolutions $\bar{f}^{*n}(S)$. Consequently, even if high expansion orders of $\psi(\sigma)$ are required, a finite expansion order of $f(\sigma)$ can be sufficient, if the magnitude of the low expansion order coefficients $\bar{f}(S)$ is large enough.

\textit{Remark.---} Due to $\psi(\sigma )= \exp(f_{A}+f_B+f_{AB}) = \exp(f_A)\exp(f_B)\exp(f_{AB})$, with $f_r = \sum_{S \subseteq r} \bar{f}(S)\mathcal{X}_S(\sigma_S)$ only coefficients $\bar{f}(S_A \cup S_B)$ with $S_A,  S_B \neq \emptyset$ can contribute to the entanglement between subsystem $A$ and $B$ of the quantum state.

\section{Deriving entanglement bounds}
In this section, we derive the entanglement scalings for an NQS that contains expansion orders up to order $n \leq 1/2 X$, with X being the smaller bipartition of a system. The binary entropy approximation is given by~\cite{Cover_2005,galvin_2014}:
\begin{equation}
    S_\mathrm{bound}(n,X) 
    = - n \ln \left(\frac{n}{X} \right) - (X-n) \ln \left(1  - \frac{n}{X} \right).
\end{equation}
Of interest are the three distinct regimes: constant expansion order, area law scaling, and volume law scaling.

\textit{Constant expansion order $n$.---}
In the large $X$ regime, a constant expansion order $n$ yields maximally logarithmic entanglement growth:
\begin{equation}
\label{App:Sbound}
    S_\mathrm{bound}(n,X) 
    = \underbrace{- n \ln \left(\frac{n}{X} \right)}_{\propto n \ln(X)} - \underbrace{(X-n) \ln \left(1  - \frac{n}{X} \right)}_{\propto n}.
\end{equation}

\textit{States with area law entanglement.---}
\label{App:Entanglementscalings}
To simplify the derivation, we use the Ansatz $n = \sqrt{X}f(X)$:
\begin{equation}
\begin{split}
    &S_\mathrm{bound}(n,X)/\sqrt{X} = c \\
    &= f(x) \ln \left(\frac{\sqrt{X}}{f(X)} \right) -(\sqrt{X}-f(X)) \underbrace{\ln \left(1  - \frac{f(X)}{\sqrt{X}} \right)}_{\overset{f(X) \ll \sqrt{X}}{\approx} - \frac{f(X)}{\sqrt{X}}}.
\end{split}
\end{equation}
In the case $f(x) = const.$ and $f(X) \ll \sqrt{X}$, we get a scaling of $S_\mathrm{bound} \propto \sqrt{X} \ln \left( X \right)$. Therefore, we can assume $\partial_X f(X)<0$ and $f(X) \ll \sqrt{x}$ in the large X limit. With these assumptions, the above equation reduces to:
\begin{equation}
    c\approx f(X) \left( \ln \left(\sqrt{X}\right) -\ln \left(f(X)\right) \right) - \underbrace{(f(X)-\frac{f^2(X)}{\sqrt{X}})}_{\longrightarrow 0}.
\end{equation}
This leaves us with three cases to consider:
\begin{itemize}
    \item $f(X) \propto \frac{1}{\ln(X)}$: this cancels the scaling of $\ln\sqrt{X}$ and $-\ln(f(X)) \propto \ln(X)$ and therefore solves the equation. 
    \item $f(X)$ decays faster than $\frac{1}{\ln(X)}$:
    $f(X)\ln \left(f(X)\right)$ becomes the leading term, but decays to zero.
    \item $f(X)$ decays slower than $\frac{1}{\ln(X)}$: the leading term $f(X)\ln \left(\sqrt{X}\right)$ increases in magnitude with X. This includes more than pure area law scaling.
\end{itemize}
Consequently, CQS need a correlation scaling of $n \propto \frac{\sqrt{X}}{\ln(X)}$ to be able to represent states with area law entanglement.

\textit{States with volume law entanglement.---} We start again with $n=X f(X)$:
\begin{equation}
c =  -f(X)\ln \left(f(X) \right) - (1-f(X))\ln \left(1  - f(X) \right),
\end{equation}
which is the binary entropy function $H(f(X)) = c$. The solution $f(X) = H^{-1}(c)$ has one valid, X independent solution for $f(X) \leq 1/2$ (the regime where our approximation holds). Thus, n has to scale with the volume $X$ to obtain volume law entanglement scaling.

\bibliography{biblio.bib}

\end{document}